\documentclass[aps,prx,floatfix,nofootinbib,twocolumn,superscriptaddress,longbibliography]{revtex4-2}
\usepackage{dsfont}

\usepackage{amsmath}
\usepackage{amsfonts}
\usepackage{physics}
\usepackage{amssymb}
\usepackage{graphicx}
\usepackage{color}
\usepackage{accents}
\usepackage{bbold}
\usepackage[colorlinks=true, citecolor=blue]{hyperref}
\usepackage[normalem]{ulem}

\makeatletter 
\newsavebox{\@brx}
\newcommand{\llangle}[1][]{\savebox{\@brx}{\(\m@th{#1\langle}\)}%
  \mathopen{\copy\@brx\kern-0.5\wd\@brx\usebox{\@brx}}}
\newcommand{\rrangle}[1][]{\savebox{\@brx}{\(\m@th{#1\rangle}\)}%
  \mathclose{\copy\@brx\kern-0.5\wd\@brx\usebox{\@brx}}}
\makeatother

\newlength{\dhatheight} 

\newcommand{\qed}{\nobreak \ifvmode \relax \else
      \ifdim\lastskip<1.5em \hskip-\lastskip
      \hskip1.5em plus0em minus0.5em \fi \nobreak
      \vrule height0.75em width0.5em depth0.25em\fi}
      
\newcommand\varpm{\mathbin{\vcenter{\hbox{%
  \oalign{\hfil$\scriptstyle+$\hfil\cr
          \noalign{\kern-.3ex}
          $\scriptscriptstyle({-})$\cr}%
}}}}
\newcommand\varmp{\mathbin{\vcenter{\hbox{%
  \oalign{$\scriptstyle({+})$\cr
          \noalign{\kern-.3ex}
          \hfil$\scriptscriptstyle-$\hfil\cr}%
}}}}

\long\def\/*#1*/{}

\begin{document}
\title{Protocol Discovery for the Quantum Control of Majoranas by Differentiable Programming and Natural Evolution Strategies}

\author{Luuk Coopmans}
\thanks{Co-first authors.}
\affiliation{Dublin Institute for Advanced  Studies, School of Theoretical  Physics, 10 Burlington Rd, Dublin, Ireland.}
\affiliation{School of Physics, Trinity College Dublin, College Green, Dublin 2, Ireland}
\author{Di Luo}
\thanks{Co-first authors.}
\affiliation{Department of Physics and IQUIST and Institute for Condensed Matter Theory,  University of Illinois at Urbana-Champaign, IL 61801, USA} 
\author{Graham Kells}
\affiliation{Dublin Institute for Advanced  Studies, School of Theoretical  Physics, 10 Burlington Rd, Dublin, Ireland.}
\author{Bryan K. Clark}
\affiliation{Department of Physics and IQUIST and Institute for Condensed Matter Theory,  University of Illinois at Urbana-Champaign, IL 61801, USA} 
\author{Juan Carrasquilla}
\affiliation{Vector Institute for Artificial Intelligence, MaRS Centre, Toronto, Ontario, Canada}
\affiliation{Department of Physics and Astronomy, University of Waterloo, Ontario, N2L 3G1, Canada}

\date{\today}
\begin{abstract}
Quantum control, which refers to the active manipulation of physical systems described by the laws of quantum mechanics, constitutes an essential ingredient for the development of quantum technology. 
Here we apply Differentiable Programming (DP) and Natural Evolution Strategies (NES) to the optimal transport of Majorana zero modes in superconducting nanowires, a key element to the success of Majorana-based topological quantum computation. We formulate the motion control of Majorana zero modes as an optimization problem for which we propose a new categorization of four different regimes with respect to the critical velocity of the system and the total transport time. In addition to correctly recovering the anticipated smooth protocols in the adiabatic regime, our algorithms uncover efficient but strikingly counter-intuitive motion strategies in the non-adiabatic regime.
The emergent picture reveals a simple but high fidelity strategy that makes use of  pulse-like jumps at the beginning and the end of the protocol with a period of constant velocity in between the jumps, which we dub the jump-move-jump protocol. We provide a transparent semi-analytical picture, which uses the sudden approximation and a reformulation of the Majorana motion in a moving frame, to illuminate the key characteristics of the jump-move-jump control strategy. We verify that the jump-move-jump protocol remains robust against the presence of interactions or disorder, and corroborate its high efficacy on a realistic proximity coupled nanowire model. Our results demonstrate that machine learning for quantum control can be applied efficiently to quantum many-body dynamical systems with performance levels that make it relevant to the realization of large-scale quantum technology. 
\end{abstract}
\maketitle

\section{Introduction}

Through the use of a wide array of promising experimental platforms ranging from superconducting qubits \cite{Gambetta2017} to trapped ions \cite{Cirac1995,Bohnet2016,Bruzewicz2019}, optical lattices \cite{Gross2017} and nitrogen-vacancy centers \cite{Zhou2017,Casola2018}, scientists are exploring ground-breaking ways to build quantum technology with an eye on deepening our understanding of complex natural systems, improving artificial intelligence, and impacting industry more broadly. While promising, several fundamental and practical difficulties must be overcome for quantum machines to become practical~\cite{DiVincenzo2000}. Quantum control, which studies the manipulation of physical systems whose behaviour is dominated by the laws of quantum mechanics, remains a fundamental ingredient in the quest for practical quantum technology. Thus, the development of principles and strategies for quantum control \cite{Baksic2016,Rabitz824, PhysRevLett.114.170501, RevModPhys.91.045001, delCampo2015} remains an essential task for future quantum technologies~\cite{macfarlane2003}, e.g., in the preparation of complex quantum states in quantum computing and quantum simulators.  
 
The practical encoding and manipulation of quantum information has however been hampered by the presence of noise and decoherence inherent to brittle quantum devices. Majorana zero modes, which are special zero-energy quasi-particles with non-abelian braiding statistics that can potentially be realized in proximity-coupled superconductors~\cite{PhysRevLett.100.096407,PhysRevLett.105.077001,PhysRevLett.105.177002,Alicea2011,Kjaergaard2012,Leijnse2012,Mourik2012,Deng2012,Das2012,Lutchyn2018}, represent a promising alternative approach due to their non-local nature and anticipated topological protection against local errors~\cite{Kitaev2001,Kitaev2003,Sarma2015,Nayak2008,Stanescu2013,Aasen2016}. The key to manipulating this topologically encoded quantum information is the development of protocols to transport the Majoranas, as will be necessary for their braiding and measurement. While the simplest approaches proposed to move the Majoranas adiabatically, decoherence processes such as quasiparticle poisoning \cite{Rainis2012}, motivate the need for faster and more efficient control protocols.

  \begin{figure*}[t!]
    \centering
    \includegraphics[width=1.0\linewidth]{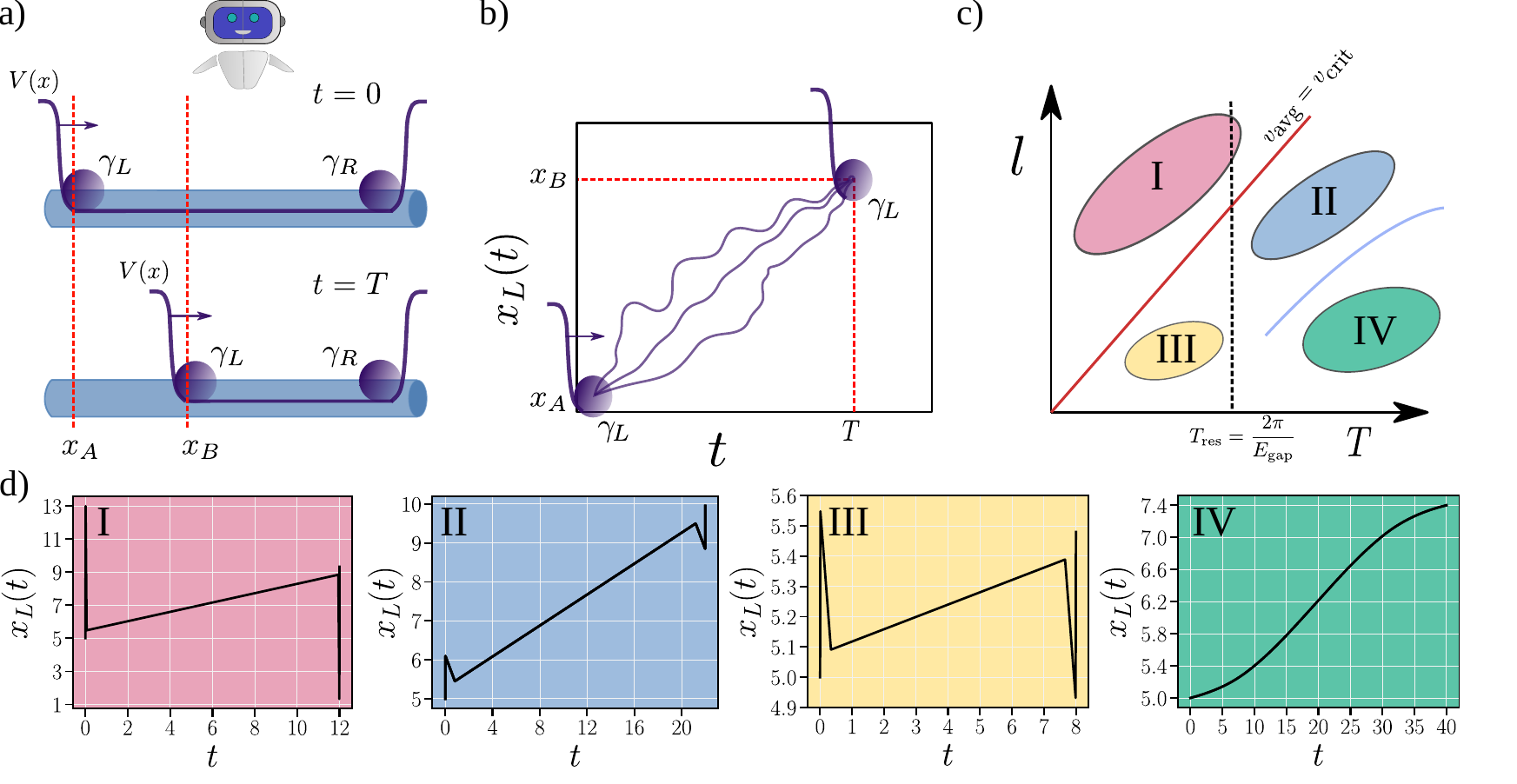}
    \caption{a) In the Majorana Game the agent moves the left Majorana $\gamma_L$ from a position $x_A$ to a position $x_B$ by tuning the external potential profile $V(x,t)$. b) The agent attempts different movement paths and tries to find the optimal one that minimizes the infidelity $\mathcal{I}(T)$. c) Different Majorana control regimes and their corresponding strategies (boxes in d)): I) Critical regime ($v_{\text{avg}}>v_{\text{crit}}$), II) sub-critical regime ( $v_{\text{crit}}/2 < v_{\text{avg}} < v_{\text{crit}}$), III) short time $T$, low velocity regime and IV) (super)-adiabatic regime (long time $T$  and low velocity $v_{\text{avg}}$). d) In regimes I-II-III we find the jump-move-jump (JMJ) strategy whereas in regime IV we recover a smooth super-adiabatic protocol. The infidelities are $\mathcal{I}(T)=0.3575$ in regime I, $\mathcal{I}(T)=0.1589$ in II, $\mathcal{I}(T)=0.0057$ in III and $\mathcal{I}(T)=0.0005$ in regime IV; the parameters for the JMJ strategies displayed in regimes I, II, III can be found in App. \ref{app:model} and we have set $x_L(0)=x_A=5.0$ to ensure a smooth potential profile $V(x,t)$.}  
    \label{fig:majorana_game_setup}
\end{figure*}

In this context, machine learning (ML) offers a powerful and unifying approach to the study, design, benchmarking, and control of quantum systems and devices. Motivated by a series of remarkable technological breakthroughs in research areas as diverse as computer vision~\cite{voulodimos2018}, natural language processing~\cite{young2018} and genomics~\cite{eraslan2019}, physicists have started to explore the potential of ML for fundamental research~\cite{RevModPhys.91.045002}. In particular, researchers interested in quantum many-body physics have initiated the development of a machine learning perspective on the many-body problem~\cite{carrasquilla2020} with recent advances such as neural network representation of quantum  states~\cite{lagaris1997,carrasquilla2017,carleo2017,kochkov2018variational,PhysRevLett.122.226401,ferminet,paulinet,autoregressive_wf,rnn_wf,topo_wf,Gao2017,Glasser_2018}, neural network tomography~\cite{torlai2018,torlai2018a,carrasquilla2019}, machine learning classification of phases~\cite{carrasquilla2017}, as well as advances in quantum chemistry~\cite{PhysRevLett.108.058301,doi:10.1002/qua.24954,schutt2019}, 
 density functional theory~\cite{PhysRevLett.108.253002,doi:10.1002/qua.25040} and the acceleration of Monte Carlo simulations~\cite{PhysRevB.95.041101,PhysRevB.95.035105,PhysRevE.100.043301}, preparation of quantum states~\cite{melnikovActiveLearningMachine2018,yaoReinforcementLearningManyBody2020}, quantum feedback~\cite{PhysRevX.8.031084}, 
among many others~\cite{dunjko2018,carrasquilla2020}.
ML applications in the context of 
quantum control~\cite{PhysRevX.8.031086,niu2019,zhangWhenDoesReinforcement2019,Schafer2020} are largely based on reinforcement learning (RL) techniques, which are the key ingredient behind major artificial intelligence breakthroughs in game play~\cite{mnih2015,silver2016}. A RL perspective on quantum control has been developed in the context of quantum state preparation~\cite{PhysRevX.8.031086}, where the authors reexamine the quantum state preparation in terms of a modified version of the Watkins Q-learning algorithm~\cite{10.5555/3312046}. Likewise, Ref.~\cite{niu2019} considers the problem of gate synthesis from a RL viewpoint.
 
In this work we exploit Differentiable Programming (DP)~\cite{Baydin2017} and Natural Evolutionary Strategies (NES)~\cite{rechenberg1973} to find optimal strategies to transport Majorana zero modes. We reformulate the complex task of transporting a Majorana zero mode as a quantum control optimization problem amenable to advanced ML techniques. This can be most easily understood from a reinforcement learning perspective where the movement of the Majoranas can be recast as a ``game'', see Fig. \ref{fig:majorana_game_setup}. In this game the player (agent) has to move a Majorana from a position $x_A$ to a position $x_B$ in a fixed amount of time $T$. At the end of the game the agent gets rewarded depending on how well the Majorana reached its target state at position $x_B$. The objective of the agent is to learn the best strategy (path of the Majorana) that maximizes the reward. 

Our machine learning techniques discover a novel and counter-intuitive approach to transporting Majorana zero modes, here dubbed the {\it jump-move-jump} approach and exemplified in Fig.~\ref{fig:majorana_game_setup}, which yields a high fidelity control significantly superior in terms of speed and quality to previous strategies applied to this problem~\cite{Scheurer2013,Karzig2015,Conlon2019}. The jump-move-jump protocol, which is relevant for transporting Majoranas over a large distance $l$ in a short time $T$,  makes use of pulse-like jumps at the beginning and end of the protocol (see regions I-III of Fig.~\ref{fig:majorana_game_setup}), with a period of nearly constant velocity between the jumps. This is particularly interesting given the previous results for other models showed that bang-bang protocols~\cite{Karzig2015} were the most efficient approach to this form of quantum control.

Using the insight gleaned from our ML-inspired control protocols, we construct the core strategy for these paths and provide a theoretical understanding for the protocol by analyzing the motion of the Majoranas in a moving frame. We find that these protocols simultaneously use the stability of the system at finite velocities together with the fact that small sudden jumps in wall position do not significantly reduce the ground-state fidelity.  In contrast, in the limit where there is a significant total time $T$ to move a relatively small distance $l$, the best approach is to follow smooth protocols that follow an adiabatic path. Our ML technology recovers these protocols without any prior knowledge (see region IV of Fig.~\ref{fig:majorana_game_setup}). In addition,  we verify that the jump-move-jump protocol retains its usefulness in the presence of interactions or disorder, and also corroborate its high efficacy using the more experimentally relevant proximity-coupled semiconductor nanowire model. 

We have structured this paper as follows: in section \ref{sec:Setup} we introduce the setup of the Majorana control problem. In section \ref{sec:MLmethods} we introduce the optimization methods DP and NES for which we provide the optimization results in section \ref{sec:results} which we benchmark against a standard simulated annealing method. In section \ref{sec:results} we also highlight some of the advantages of the DP and NES methods compared to others. We describe and analyse the physical mechanisms behind the jump-move-jump strategy in section \ref{sec:JMJ}. Then we provide some numerical results of the robustness of the jump-move-jump strategy with respect to interactions or disorder in section \ref{sec:robustness} before concluding our work. 

We also include several appendices where we provide extra details regarding the moving frame in App.~\ref{app:majorana_motion}, ML methods in App.~\ref{app:gradient_analysis},~\ref{app:NES} and~\ref{app:mlqcontrolcomp}, extra results for the proximity-coupled semiconductor nanowire in App.~\ref{app:semiconductor} and the robustness of the JMJ strategy with respect to interactions and disorder in App.~\ref{app:int_dis}. Finally in App.~\ref{app:model} and App.~\ref{app:JinfCost} we discuss extra details for the analysis of the JMJ strategy. 

\section{Setup of the Majorana Control Problem}\label{sec:Setup}

To model the movement of the Majorana zero modes in superconducting nano-wires we use a one-dimensional (1D) Kitaev Chain~\cite{Kitaev2001} described by the Hamiltonian \begin{equation}
\label{eq:Kitaevchain}
	\begin{split}
	\mathcal{H}(t) &=  - \sum_{x=1}^{N} [\mu(x)-V(x,t)]  (c^\dagger_{x}c_{x }-1/2)    \\
	&- w \sum_{x=1}^{N-1} (c^\dagger_{x}c_{x+1} + h.c) +\Delta  \sum_{x=1}^{N-1}( c^\dagger_{x} c^\dagger_{x+1} + h.c. ),  
\end{split}
\end{equation} where $c^{(\dagger)}_{x}$ are fermionic annihilation (creation) operators, $\mu(x)$ the onsite chemical potential, $V(x,t)$ a time-dependent external potential, $w$ the hopping amplitude and $\Delta$ the p-wave superconducting gap. This model can be realized effectively in variety of setups by proximity coupling to a conventional s-wave superconductor~\cite{PhysRevLett.100.096407,PhysRevLett.105.077001,PhysRevLett.105.177002,Brouwer2011,Chung2011,Kjaergaard2012,Leijnse2012,Perge2013}.  
When $\abs{\mu}\geq\abs{2w}$ the gap in the energy spectrum closes inducing a  transition to a topological trivial phase~\cite{Kitaev2001,Bernevig2013}. For an open chain, a pair of Majorana zero-modes are found to reside on the edges of the wire if the bulk is topologically non-trivial. The existence of such Majorana modes implies the existence of a degenerate space of ground states. The ground state of the system is protected by a robust topological energy gap, which in the continuum limit of the model is given by $ \min[ \mu_c, \Delta_c k_F]$. As a consequence of this protection, two pairs of Majorana zero-modes can be used to produce a qubit whose information content is protected non-locally~\cite{Sarma2015}.

Manipulation of the information encoded in the ground state requires the braiding of the Majorana quasi-particles \cite{Read2000,Ivanov2001,Kitaev2003,Stern2004,Stone2006,Nayak2008,Alicea2011, Stanescu2013}, while staying as much as possible within the degenerate ground-state space. To achieve this, a generic strategy is to try to perform the braiding adiabatically. An adiabatic path generically must have a total time $T$ larger then the inverse gap; in practice the size of the topological gap makes such a strategy prohibitively slow in view of numerous sources of decoherence in proximity coupled setups~\cite{Rainis2012,Budich2012,Konschelle2013,Ng2015,Pedrocchi2015,Ying2015,Knapp2016,Knapp2018,Huang2019}. 
 
In light of this, our strategy in what follows (see also \cite{Karzig2013,Scheurer2013,Conlon2019}), is to attempt to move Majoranas as quickly and efficiently as possible. To this end, and to have smooth control over the position of the Majorana bound states on the lattice, we encode the external potential profile as
\begin{equation}
\label{eq:Potentialprofile}
V(x,t) = V_{\text{height}} [ f(x-x_L(t))+f(x_R-x)]
\end{equation} where $V_{\text{height}}$ is the maximum height of the outer potential walls and $f(x)=1/(1+\text{exp}(x/\sigma))$ is a sigmoid function shifted by the wall positions $x_{L(R)}$ (Fig. \ref{fig:majorana_game_setup} a). The Majoranas are localized at the outer edges of the potential profile, which can be seen as domain walls between topological and non-topological phases since $V_{\text{height}} \gg 2w$. The position of these domain walls can be be controlled through $x_{L(R)}$; to move the left Majorana we give the agent control over $x_L(t)$ as a function of time.  

The motivation for this specific form of potential profile comes from experimental proposals for moving the Majorana by the so called piano-key potentials~\cite{Alicea2011,Scheurer2013,Bauer2018}. In this proposal the position of the domain wall in the wire is controlled by applying gate electrodes. \footnote{For a more detailed discussion of these effects in more complicated geometries see Refs. \cite{PhysRevX.8.031041, PhysRevB.98.035428}. For a discussion on alternative braiding schemes see  \ref{sect:OtherSchemes}.} At variance with protocols found in~\cite{Karzig2015}, the presence of non-linearities in the potential profile implies that bang-bang protocols, which are expected for linear control \cite{Yang2017,Brady2020}, may no longer be optimal. As a consequence of this, it is conceivable that the space of possible solutions is expanded in our setting.

The Majorana game starts at $t=0$ with the system in the groundstate of Eq. \ref{eq:Kitaevchain} with $x_L(0)=x_A$. The aim is to reach the groundstate $\ket{\Psi_B}$ with $x_L(T)=x_B$ located at a distance $x_B-x_A=l$ spending a total time $T$. A natural choice to quantify the accuracy of the protocol is the infidelity 
\begin{equation}\label{eq:infidelity}
    \mathcal{I}(T) = 1-|\bra{\psi_B}\mathcal{T}e^{-i\int_0^T\mathcal{H}(t)dt}\ket{\psi_A}|^2 \equiv  1-\mathcal{F}(T).
\end{equation}
 Here, $\mathcal{H}(t)$ is the time-dependent Hamiltonian that describes the Majorana wire setup Eq. \ref{eq:Kitaevchain} during the protocol. Whereas $\mathcal{I}(T)=0$ means that we have fully preserved the encoded quantum information,  $\mathcal{I}(T)=1$ implies that information has been completely lost.

A key timescale related to the control problem is $T_{\text{res}}=\frac{2\pi}{\Delta k_F}$, which naturally arises from the response of the system to boundary wall oscillations (see ~\cite{Conlon2019} and App. \ref{app:majorana_motion}). $T_{\text{res}}$ corresponds to the time above which one can make changes slowly enough for genuinely super-adiabatic motion \cite{Berry1987, Deschamps2008,Scheurer2013}. Super-adiabaticity in this scenario allows the static ground state to be accelerated into the ground state of a moving frame, provided the transition is done slowly enough, i.e., in times large compared to $T_{\text{res}}$. Another important additional physical constraint is the presence of a critical velocity $v_{\text{crit}}=\Delta$~\cite{Karzig2013,Scheurer2013} above which the moving frame Hamiltonian becomes gapless (App. \ref{app:majorana_motion}) and the ground states lose their topological protection. 
 
Based on this we divide up the Majorana control problem into four different regimes (I-IV) (see Fig.~\ref{fig:majorana_game_setup}~c).

\begin{itemize}
    \item Regime I corresponds to the critical regime in which the Majorana must move on average $v_{\text{avg}}=l/T$ above the critical velocity. In this regime the ground state fidelity is expected to rapidly decrease to zero.

\item Regime II is the sub-critical regime for which the velocity is on average close to but nonetheless below $v_{\text{crit}}$. This regime is open ended in both time $T$ and length $l$. The key feature distinguishing this regime from regime IV below is character of the found optimal protocols.

\item Regime III is again a sub-critical regime, defined for the times that are short with respect to the resonance time $T_{\text{res}}$ but also has low velocity. This region cannot be used to efficiently move Majorana states over long distances, but we expect it to be relevant for braiding protocols based on small relative movements that change the effective couplings between Majoranas. 

\item Finally regime IV is the adiabatic regime in which we are above $T_{\text{res}}$ and we have sufficient time to expect that slow ramp-up/ramp-down  protocols~\cite{Scheurer2013, Conlon2019} from earlier studies to be optimal.  Ideally one would always like to be in this regime, however a gradual build up of noise and decoherence may make it necessary to get things done more quickly.   
\end{itemize}

\subsection{Relevance to other braiding schemes}
\label{sect:OtherSchemes} 
It is worth mentioning here that the aforementioned trade-off between adiabaticity and the need to perform operations quickly has led to the emergence of other approaches to braiding.  These schemes seek to circumvent the need to manipulate/shuttle Majorana's over excessively long distances.  This includes local \cite{Alicea2011,Sau2011,Halperin2012} and non-local \cite{vanHeck2012,Burrello2013} Majorana coupling in wire-networks,  and the implementation of one-way computation schemes \cite{Bonderson2008,Bonderson2013} via projective charge measurements \cite{Aasen2016,Vijay2016,Karzig2017,Plugge2017,Knapp2020,Zeng2020}.  

All of these methods still require some level of adiabatic control, or some effective short-cutting thereof (see e.g. ~\cite{torrontegui2013,PhysRevB.91.201102}) and there can also be some additional downsides. The interaction/coupling schemes for example require precise control of the couplings between neighbouring Majoranas and the loss of some topological protection \cite{Kells2014,Pedrocchi2015}.  In implementations of wire measurement-only schemes one must also tune a coupling between wire and a quantum dot \cite{Aasen2016,Plugge2017} and accept state manipulation that is inherently probabilistic \cite{Zeng2020}.

The phase diagram that we have outlined above is directly applicable to the local coupling schemes, where the relative Majorana position can be seen as a proxy for the coupling strength. A similar proxy may also hold for measurement-only schemes via the distance between topological boundary and quantum dot, although the connection here is harder to make directly because one should also account for fermion number conservation \cite{Aasen2016, Vijay2016, Karzig2017, Plugge2017,Knapp2020}.  That said, the DP and NES methods we apply below do have direct relevance for even more sophisticated numerical treatments of this problem.  We refer to our concluding remarks for further discussion of this point.

\section{Machine Learning Methods}\label{sec:MLmethods}
In our study we apply two machine learning strategies, namely Differentiable Programming (DP) and Natural Evolution Strategies (NES), to minimize the infidelity Eq. \ref{eq:infidelity} with respect to the position of the domain wall $x_\text{L}(t)$. \footnote{We note that the control can also be represented by the velocity or a neural network parameterization of the domain wall from which the position $x_L(t)$ can be extracted.} In this section we introduce these methods in a way that directly applies to our Majorana control setup. For the interested reader we provide the corresponding programming codes in \cite{Github}. 

\subsection{Differentiable Programming}
DP is a programming paradigm to evaluate gradients of computer programs~\cite{Wengert1964} that 
has been recently introduced as an optimization method in physics applications like Tensor Networks~\cite{Liao2019,Xie2020}, Monte Carlo~\cite{Sorella2010,kochkov2018variational,zhang2019automatic}, and optimal control~\cite{Leung2017,Schafer2020}. DP obtains numerically exact gradients with similar computational time complexity as the forward evaluation of the computer program due to the use of backward propagation~\cite{Rumelhart1986, BAUR1983317,Morgenstern1985}.

In our Majorana control optimization problem we are interested in the total derivative of the infidelity with respect to the control $\frac{d\mathcal{I}(T)}{d x_L(t)}$. As shown in App. \ref{app:gradient_analysis}, the algorithm to compute $\mathcal{I}(T)$ consists of a sequence of elementary operations $f_i$, in DP language known as primitives, which maps the input control $x_\text{L}(t)$ to the value of the infidelity by~$\mathcal{I}(T)=f_n \circ f_{n-1} \circ \cdots \circ f_1(x_L(t))$. Moreover, the derivatives of each individual operation $f_i$ are known and the total derivative $\frac{d\mathcal{I}(T)}{d x_L(t)}$ can be assembled by recursively applying the chain rule with automatic differentiation (see App. \ref{app:gradient_analysis}). 

In practice, we write a code to evaluate the real-time dynamics of the Majorana induced by the time-dependent profile $V(x,t)$ such that all the individual operations are differentiable. To obtain the necessary gradients we use a language that supports automatic differentiation, which in our case is the JAX library \cite{jax2018github}. The gradients are used to minimize the infidelity of the final state using Gradient Descent (GD) and ADAM~\cite{Adam}. For GD  this is done iteratively, where at each iteration of the algorithm the protocol is updated as $x_L(t) \rightarrow x_L(t)-\nabla_{x_L(t)}I(T)$. Moreover since neural networks consist of differentiable elementary operations, we also explore representing and training the control $x_L(t)$ as the output of a neural network for which the weights are updated with the standard update scheme ADAM.

\subsection{Natural Evolution Strategies}

Evolution Strategies is a family of black-box optimization algorithms inspired by natural evolution~\cite{rechenberg1973,h.-p.schwefel1977numerische-opti}. These methodologies have been recently revisited in the context of machine learning~\cite{JMLR:v15:wierstra14a}, in particular in reinforcement learning~\cite{salimans2017}, where OpenAI demonstrated that a particular incarnation of the algorithm called natural evolution strategies (NES)  offers a powerful alternative to popular Markov-decision process-based reinforcement learning methods~\cite{Sutton1998}.

NES starts with an objective function $F(\phi)$ that acts on parameters $\phi$ from a population described by a distribution $p_{\theta}$, where $\theta$ parameterizes the population distribution. In our work, we choose the objective function $F(\phi)$ to be the infidelity $\mathcal{I}(T,\phi)$ in Eq.~\ref{eq:infidelity}. Depending on the setting, the parameter $\phi$ corresponds to either the position profile $x_L(t)$, the velocity $v_L(t)=\dot{x}_L(t)$ or the parameters of a neural network whose output is $x_L(t)= \text{NN}_{\phi}(t)$.    

The NES algorithm proceeds to optimize the expectation value of the objective function $\mathbb{E}_{\phi \sim p_{\theta}}\left[ \mathcal{I}(T,\phi)\right]$ over the population. We choose $p_{\theta}$ to be a Gaussian distribution with mean $\theta$ and diagonal covariance matrix $\Sigma = \sigma^2 I$ with $\sigma=0.1$, i.e., $\mathcal{N}(\theta, \sigma^2 I)$. It follows that the gradient of the cost function is (see App.~\ref{app:NES})  
\begin{equation}
\begin{split}
     \nabla_{\theta} \mathbb{E}_{\phi \sim \mathcal{N}(\theta, \sigma^2 I)}\left[ \mathcal{I}(T,\phi)\right] 
    &= \nabla_{\theta} \mathbb{E}_{\epsilon \sim \mathcal{N}(0,I)} \left[ \mathcal{I}(T,\theta + \sigma \epsilon) \right]\\
    &= \mathbb{E}_{\epsilon \sim \mathcal{N}(0,I)} \left[ \mathcal{I}(T,\theta + \sigma \epsilon) \epsilon / \sigma\right].
\end{split}
\label{eq:NES}
\end{equation}

This equation provides an efficient way for computing gradients without differentiation, but instead through the expectation of the objective function. Notice that $\mathbb{E}_{\epsilon \sim \mathcal{N}(0,I)} [ \mathcal{I}(T,\theta) \epsilon / \sigma] = 0$, which implies the above equation is equivalent to $\mathbb{E}_{\epsilon \sim \mathcal{N}(0,I)} [ (I(T,\theta + \sigma \epsilon) - I(T,\theta)) \epsilon / \sigma]$. This means that the estimation of the gradient can be seen as the finite difference of the objective function with random search \cite{salimans2017}. To update the parameters $\theta$, we apply gradient descent to $\theta$ with Eq.~\ref{eq:NES}.

\begin{figure*}[t!]	
	\includegraphics[width=1.0\textwidth]{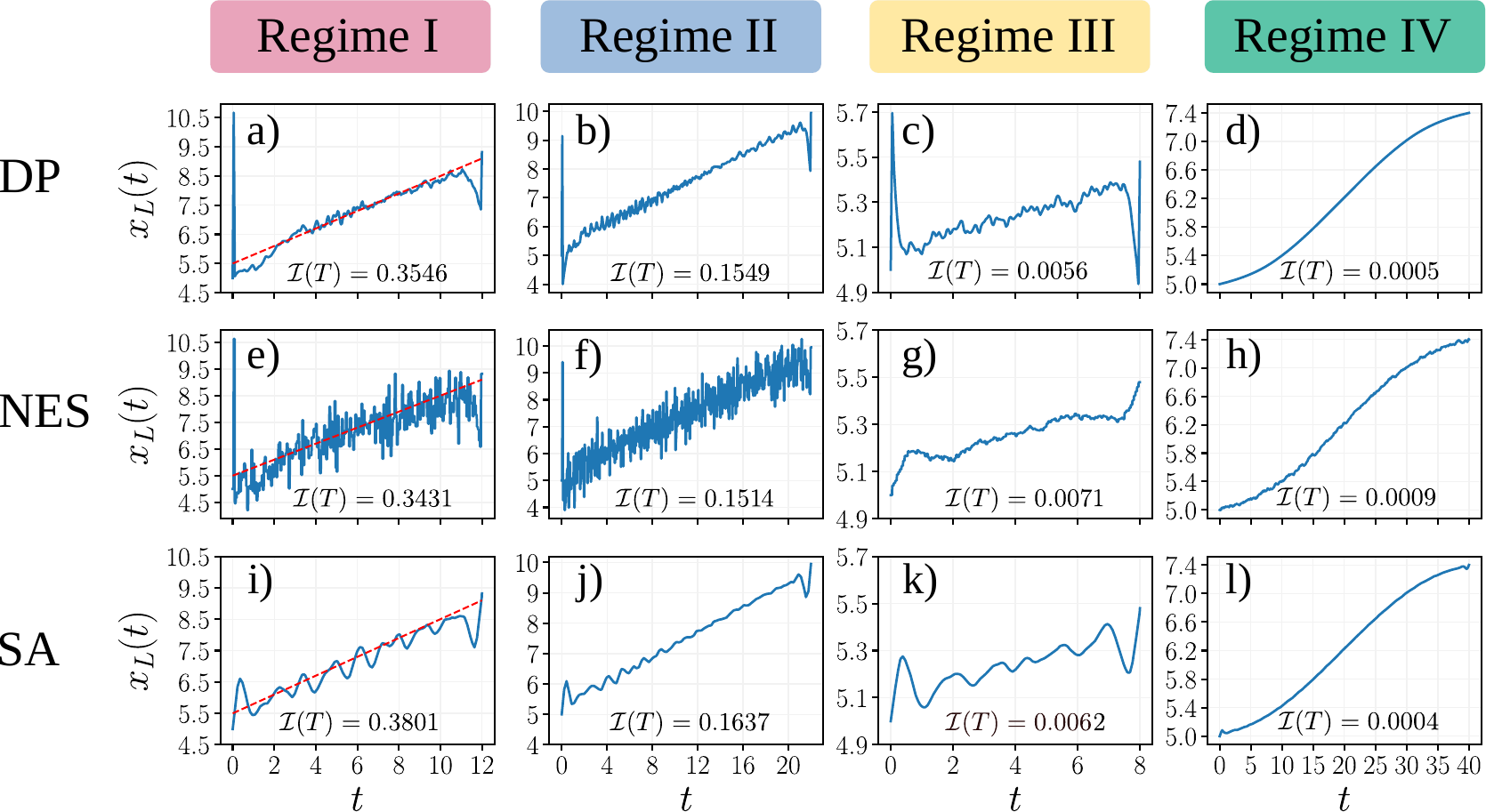}
	\caption{Overview of the optimal controls in the four different Majorana motion regimes (I-IV) obtained after optimization with Differentiable Programming (DP), Natural Evolutionary Strategies (NES) and Simulated Annealing (SA). In all panels we plot the domain wall position $x_L(t)$ as a function of time starting from $x_L(0)=x_A=5.0$. The red dotted line in regime I has a slope equal to the critical velocity. Regimes I-III show a pulse like motion at the edges and an on average constant velocity in the bulk of the protocol. In regime IV the strategy is to slowly build up to constant speed and then slow down again. In these simulations we set $N=110$, $\mu=1$, $w=1$, $\Delta=0.3$, $V_{\text{height}}=30.1$, and $\sigma=1$. The constraint parameters $(l,T)$ are given by  $\{(4.32,12),(4.95,22),(0.48,8),(2.4,40)\}$ for regimes I, II, III, and IV, respectively.}
		\label{fig:overview} 
\end{figure*}

\section{Machine Learned Strategies for Majorana control} \label{sec:results}
Our main objective is to investigate the use of DP and NES for the motion of the Majorana modes in the four different movement regimes I-IV.
We choose four different points $(l,T)$ (which fixes $v_{\text{avg}}=l/T$)
each of which belongs to a different regime. For each of these points we apply the optimization algorithms to minimize the infidelity $\mathcal{I}(T)$ in  Eq.~\ref{eq:infidelity} with respect to the control of the domain wall. The control can be parameterized by either the wall position $x_{L}(t)$, the wall velocity $v_{L}(t)$ or a neural network $x_L(t)= \text{NN}_{\theta}(t)$, where $\theta$ represent the parameters in the neural network. We tested all these different parameterizations (see App. \ref{app:extra_data}) and in the following we focus on the parameterization choices that give the best performance (the lowest infidelity values). 

For the DP optimization, we first parameterize the wall position by a neural network $x_L(t)= \text{NN}_{\theta}(t)$ and use ADAM~\cite{Adam} to optimize the parameters. We then take the resulting real-space profile $x_L(t)$ and directly re-optimize it in the position representation with GD. We find that this two-step process is beneficial; while the initial NN stage allows a quick convergence to a smooth nearly-optimal control profile, the second step allows for additional fine-tuning and more abrupt changes in the protocol. The neural network $\text{NN}_{\theta}(t)$ used for these simulations consists of three layers of 100 neurons with Rectified Linear (ReLU) activation functions followed by one output neuron with a sigmoid activation function. The learning rate for the ADAM optimization algorithm is determined empirically by running a range of values and picking the one that showed the best convergence, a method which is similar to the one used in ~\cite{Adam}. The NES optimization in regimes I/II operates directly on the wall position $x_L(t)$  and in regimes III/IV it operates on the wall velocity $v_L(t)$ from which the wall position can be extracted via integration $x_L(t)=x_A+\int_{0}^t v_L(t')dt'$. The uncertainty parameter for NES can be viewed as a trade-off of exploration and exploitation~\cite{10.5555/3312046} and was determined empirically by testing a range of values. We note that in \cite{salimans2017} it was shown that NES is a robust method with respect to different hyper-parameters in different learning settings. 

The time-dependent simulations of the fermionic system are discretized over time with a small time resolution $dt=0.01$. We allow the ML algorithms control over only a coarse grained time scale $\Delta t \geq 10  dt$ such that in the continuous time limit $dt\mapsto 0$ we get a continuous protocol $x_L(t)$ and the domain wall position is not discontinuously changing every single discrete time step $dt$. Moreover to probe the susceptibility of the optimized protocols to initial conditions, we repeat the optimizations a few times for several different starting configurations to ensure that the results are independent of the initialization. 

The results for these optimizations \footnote{Here we discuss the results obtained in the Kitaev chain while in App. \ref{app:semiconductor} we show the results obtained for the more general proximity coupled semiconducting nanowire.} in the four different regimes are shown in Fig.~\ref{fig:overview}(a-h) and can be summarized as follows. In the regimes I-II-III we can identify clear similarities between all of the optimised strategies which display sudden movements (jumps) at the beginning and end, and more gradual rates of changes in the middle of the protocol. The initial sudden movements can be roughly characterised by a rapid jump forward, followed by a less abrupt backward motion. The jumps near the end of the protocol display an analogous movements in the reverse order, although these are generally less pronounced. 
The middle of the protocols  consists of an approximately constant-velocity forward-motion that is dressed to various degrees with high frequency oscillations. We observe that the velocity in the middle part of the protocols is typically lower than $v_{\text{crit}}$ even for regime I, where the the total average velocity $v_{\text{avg}}$ (including the sudden movements) is above $v_{\text{crit}}$. The size and character of the additional oscillations largely depends on the optimization strategy being used. 

The infidelity values we find in regimes I-II-III are significantly better compared to strategies like linear motion $x_L(t)=v_{\text{avg}}t$ or bang-bang as shown in App. \ref{app:extra_data}. In the critical regime I we get an infidelity of about $\mathcal{I}(T)\approx 0.35$ compared to $\mathcal{I}(T)=0.47$ for a linear protocol whereas in regime II we get $\mathcal{I}(T)\approx 0.15$ versus $\mathcal{I}(T)= 0.22$ for linear. In the low average velocity regime III we get infidelities as low as $\mathcal{I}(T)\approx 0.006$ while linear motion gives $\mathcal{I}(T)=0.012$. The infidelity value improvement in regime I is rather surprising given that the Majorana moves on average above the critical velocity. Below, we explain why the jumps in this scenario are particularly beneficial. 

In regime IV the results show a globally different strategy: in this regime the optimal protocol is to slowly accelerate the Majorana up to some finite velocity and then slowly decelerate back down to the target position. This type of protocol was discussed in earlier work ~\cite{Scheurer2013,Conlon2019} in regimes where there is enough time to accelerate to a moving frame, i.e. regime IV. We note that due to the nearly adiabatic motion in regime IV, the values of infidelity are the lowest $\mathcal{I}(T)\approx\mathcal{O}({10^{-4}})$. These infidelity values make regime IV optimal for braiding of Majoranas in an ideal experimental setup. However, due to imperfection in the devices and their coupling to a noisy environment, balancing the infidelity associated with shorter protocols with respect to the infidelity induced by dephasing in a longer time protocol makes controlling Majorana movement in faster regimes relevant for near term experiments. 

Compared to previous studies however \cite{Knapp2016,Scheurer2013,Conlon2019} we find that the starting velocity $v_L(t=0)$ of the protocols in regime IV can be finite if one allows for small but finite infidelity values $\mathcal{I}(T)\approx\mathcal{O}({10^{-4}})$. In the obtained protocols for example we have $v_L(t=0)\neq0$ followed by an approximately smooth gradual build up and down of the velocity. Moreover, the optimization techniques with which these protocols were obtained have as additional advantage that they can fine tune the protocols up to a finer level (bigger search space) than the previously studied parameterized adiabatic protocols \cite{Conlon2019}, as we show in App. \ref{app:extra_data}). This advantage of our methods might be particularly beneficial for finding optimal protocols in the presence of disorder in the wires as discussed briefly in section~\ref{effectdisorder} and App. \ref{app:int_dis}.

Note that although we have focused on the Kitaev chain model in Eq.\ref{eq:Kitaevchain} the results and conclusions discussed here remain true on the more realistic proximity coupled nanowire models, where the JMJ protocol is optimal in the non-adiabatic regime (see details of the proximity couple nanowire simulations in  App. \ref{app:semiconductor}). Additionally, the results on the proximity couple nanowire in App. \ref{app:semiconductor} are compatible with a smooth protocol in the adiabatic regime, as well as consistent with the results on the Kitaev chain in the large Zeeman field and large strong spin-orbit coupling limits.

\subsection*{Comparison between simulated annealing and DP and NES optimization}

We now benchmark the results of the DP and NES algorithms against the standard simulated annealing (SA) method following Ref.~\cite{Karzig2015} closely. In the SA method the wall velocity $v_L(t)$ of the domain wall is iteratively updated by choosing two random intervals of length $\Delta t$ of which one interval is increased by $\Delta v$ and the other is decreased by $\Delta v$. The new infidelity $\mathcal{I}_i$ is calculated for the updated velocity profile and the move is accepted with a probability $e^{-\delta \mathcal{I}/T_{SA}}$ where $\delta \mathcal{I}=\mathcal{I}_i-\mathcal{I}_{i-1}$ is the difference in the infidelity with respect to previous iteration. The annealing temperature $T_{SA}$ is slowly cooled down to zero. The results of this benchmark in each of the four regimes are shown in Fig.~\ref{fig:overview}(i-l) and are qualitatively similar to the results obtained with NES and DP (Fig.~\ref{fig:overview}(a-h)).  

In practice, we find that SA is significantly more computationally demanding than NES or DP since a typical SA simulation entails the evaluation of the many-body infidelity for each SA update step. To obtain results with comparable infidelities, the SA simulation requires a total of $10^5$ evaluations of the infidelity while for DP we only need 440 update steps each involving a single infidelity and gradient evaluation (the latter requiring similar computational complexity as the infidelity calculation);  the NES algorithm requires 50 update steps each with 100 parallel evaluations to reach convergence. 
 
This differences may be partially attributed to the fact that SA does not take advantage of any gradient signal. Beyond these practical observations, a careful scaling analysis of the convergence of these methods requires an analysis of the convexity properties of the infidelity landscape as a function of the control parameters. This can be either done analytically for simple infidelity landscapes~\cite{chakrabartiQuantumControlLandscapes2007}, but may require a numerical investigation for control problems exhibiting a glassy landscape~\cite{PhysRevLett.122.020601}. 

Overall, we highlight that DP offers a powerful tool for quantum control as long as an accurate and differentiable physical model is available. It is found that direct gradient based gradient methods are usually more stable and efficient than RL \cite{ghosh2020learning}. In contrast, NES can optimize non-differentiable and discrete control protocols, both of which remain challenging for DP. NES can be also directly applied to experimental settings as long as a suitable objective function, such as the expectation value of a Hermitian observable or the fidelity, is available. For instance, recent experiments demonstrate the teleportation of Majorana modes in a quantum simulation of a small Majorana chain~\cite{PhysRevLett.126.090502} including access to the fidelity of the protocol. This possibility makes NES a viable tool for  controlling and designing quantum simulations of Majorana modes. 

Additional discussions detailing connections, comparisons, advantages, and disadvantages of the machine learning approaches in relation to other advanced methods used in quantum control are presented in App~\ref{app:mlqcontrolcomp}. 

\section{Jump-Move-Jump [JMJ] Majorana Control Strategy} \label{sec:JMJ}

\begin{figure}[h!]	
	\includegraphics[width=1.0\linewidth]{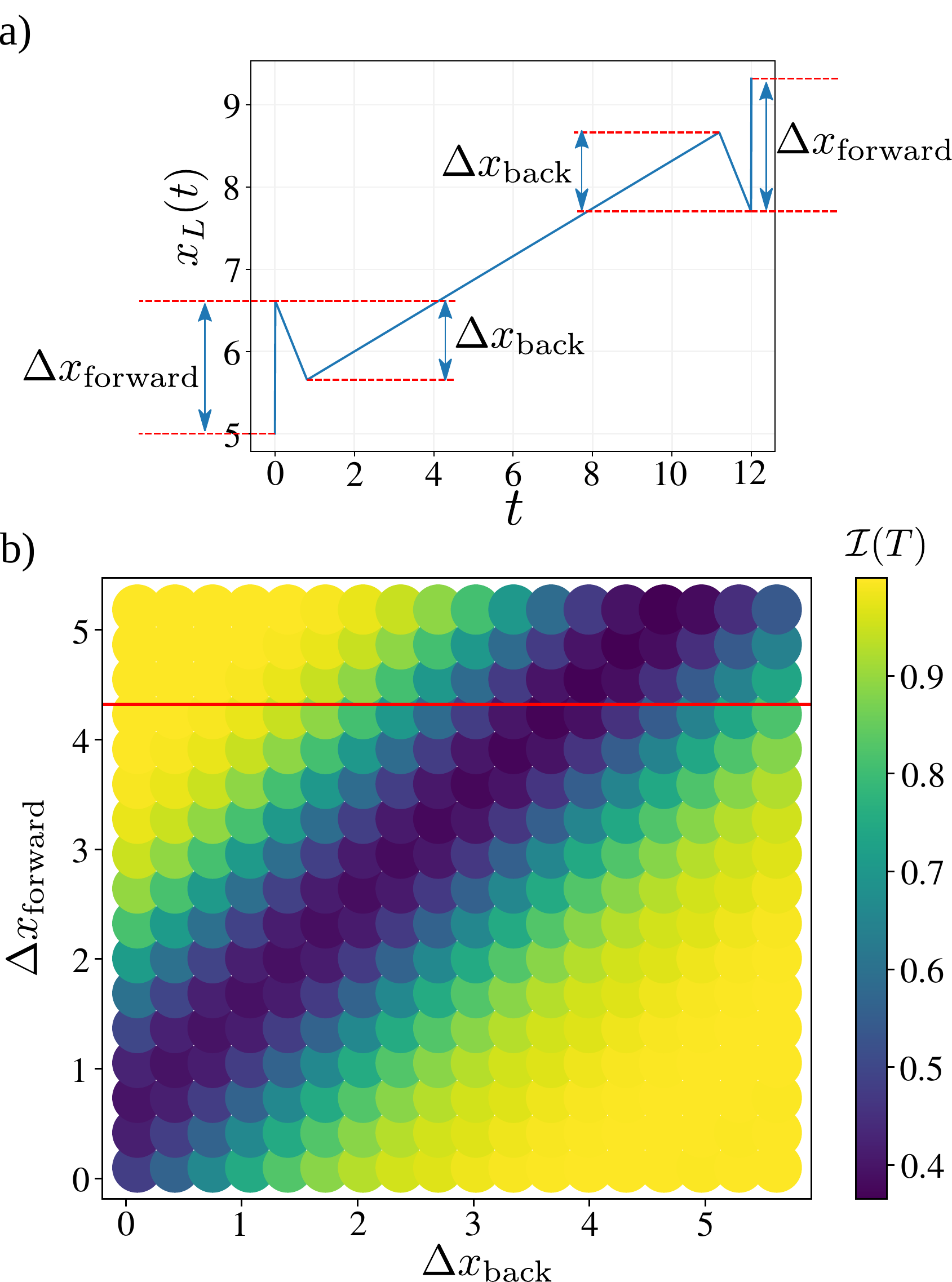}
	\caption{a) Setup of the ML-inspired simple control strategy consisting of two edge pulses with a forward jump $\Delta x_{\text{forward}}$ and backward jump $\Delta x_{\text{back}}$ together with a constant linear motion in between the dressed jumps. b) Infidelity $\mathcal{I}(T)$ surface plot for the scans over the free parameters $\Delta x_{\text{forward}}$ and $\Delta x_{\text{back}}$ of the model strategy with $\Delta t_{\text{back}}=0.07w^{-1}$ and  $\Delta t_{\text{forward}}=0.01w^{-1}$ in regime I $(l=4.32,T=12)$. The red line indicates the line where the size of the forward jump $\Delta x_{\text{forward}}$ is equal to the total movement length $l$. The parameters of the Majorana control setup are the same as in Fig.~\ref{fig:overview}. The optimal dressed jumps appear at a diagonal set of parameters $\Delta x_{\text{forward}} - \Delta x_{\text{back}} \sim C$ where $C\approx0.96$. We note that the best protocol has a jump size bigger than the movement length $\Delta x_{\text{forward}} > l$ which means that it jumps over the target position $x_L(T) = x_B$ and then jumps back within the range $x_A < x_L < x_B$ as can be seen in Fig. \ref{fig:majorana_game_setup} d) (for regimes I and III).}
		\label{fig:simple_model} 
\end{figure}

The main features of the ML strategies in regimes I, II and III can be encapsulated in a simple model strategy, the jump-move-jump strategy as shown in Fig. \ref{fig:simple_model}~ a), amenable to semi-analytical and numerical analysis. This strategy consists of two dressed jumps in the position of the domain wall at the beginning and end of the protocol interluded by a period of motion at a constant velocity $v$. The initial dressed jump comprises an instantaneous forward jump over a distance $\Delta x_{\text{forward}}$ followed by a rapid backward movement over a distance $\Delta x_{\text{back}}$ in a time $\Delta t_{\text{back}}$. Similarly, the last pulse starts with a rapid backward movement $\Delta x_{\text{back}}$ followed by a forward jump over a distance $\Delta x_{\text{forward}}$. In what follows, we assume that jumps at the beginning and end of the protocol are symmetric and are described by the same parameters. 

The simplicity of this model allows us to develop an understanding for the key mechanisms behind the ML strategies and estimate the value of the infidelity of the JMJ strategy for a wide range of parameters $l,T$ in regimes I-III (see Fig. \ref{fig:Isurface} and Fig. \ref{fig:simple_model} (b)). In section \ref{sect:xb0} below we will first argue that when we disregard the backward movements, i.e. $\Delta x_{\text{back}}=0$, the overall strategy can be explained via a trade-off (Fig. \ref{fig:Fexplain}) between the amount of fidelity loss due to the boundary jumps and the loss due to sudden changes in velocity.  

In section \ref{sect:xbn0} we also explore how this mechanism can be better controlled with the backward movements $\Delta x_{\text{back}}\neq0$ which, for a certain subset of parameters $(\Delta x_{\text{forward}}, \Delta x_{\text{back}})$, leads to a model strategy with infidelities that compare competitively with the best ML learned protocols (see the discussion in App. \ref{app:model} and Figs. \ref{fig:majorana_game_setup} (d) and \ref{fig:overview}). Our analysis shows that this dressed protocol allows one to better target the ground-state of the system in a moving (constant velocity) frame. 

\subsection{Bare Jump-Move-Jump ($\Delta x_{\text{back}} = 0$)}
\label{sect:xb0}

\begin{figure}[t!]	
	\includegraphics[width=1.0\linewidth]{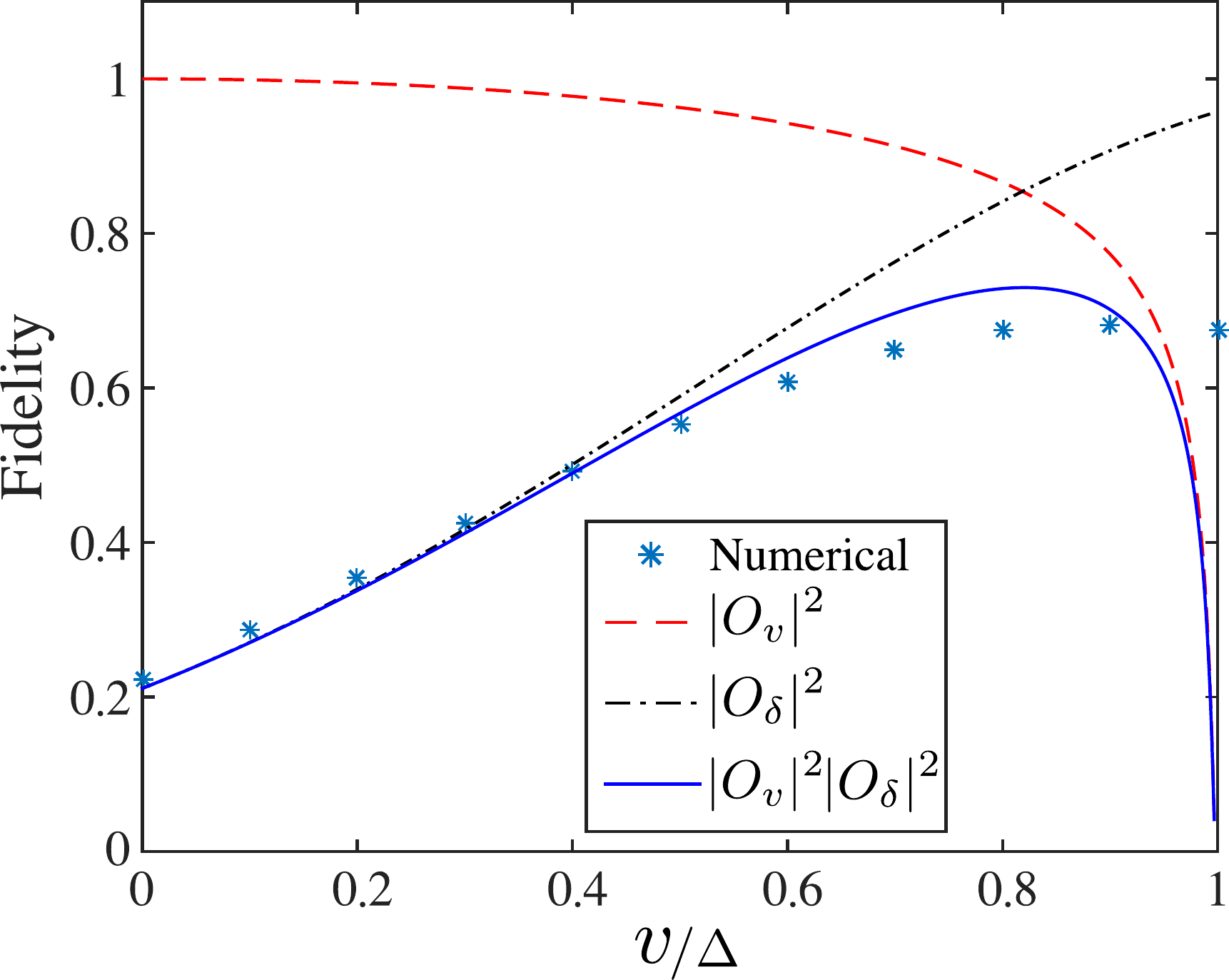}
	\caption{The jump-move-jump protocol maximizes the product of the two functions $O_\delta$ (Eq.~\ref{eq:Od}) and $O_v$  (Eq.~\ref{eq:Ov}). Here $l=4.32$ and $T=12$ (regime I) and we have used $s=2.45$ and $\beta=0.065$. The ground state projection analysis (solid blue) captures the key features of the ML-inspired (one-wall) JMJ strategy (blue stars) for low velocities $v$ but tends to overestimate the penalty for moving at velocities near $v_{\text{crit}}$.  }
		\label{fig:Fexplain} 
\end{figure}

For the case $\Delta x_{\text{back}} = 0$ we minimize the infidelity Eq. \ref{eq:infidelity} with respect to the velocity $v$ of the bulk of the protocol which fixes the instantaneous forward jump to $\Delta x_{\text{forward}} = (l -v T)/2\equiv \delta$. 
To make analytic analysis simpler we focus on the case that both the left and the right domain wall positions are being evolved simultaneously with the JMJ strategy. This means that the right wall position $x_R$ in Eq. \ref{eq:Potentialprofile} becomes time-dependent via $x_R(t)=x_L(t) + C_x$ with $C_x$ a fixed constant. This two wall scenario makes it possible to evaluate the strategy in the moving frame basis (App. \ref{app:majorana_motion}), which allows the key rationale behind the JMJ strategy to be revealed. 

To evaluate this strategy we expand the infidelity after the first jump $x_L=x_{A}\mapsto x_L=x_{A}+\delta$ in terms of the eigenbasis $\ket{\psi^i_{A+\delta}}$ of the Hamiltonian Eq. \ref{eq:Kitaevchain} with the wall at position $x_L=x_{A}+\delta$ giving 
\begin{align}
   \mathcal{I}(T) =1 - \left| \sum_i  \bra{\psi_B}  U(T) \ket{\psi^i_{A+\delta}} \braket{\psi^i_{A+\delta}}{\psi_A} \right|^2
\end{align} in which $U(T)=\mathcal{T}e^{-i\int_0^TH(t)dt}$ is the time-ordered evolution operator with $H(t)$ following the time-dependence of the strategy. We break this equation up into two separate terms
\begin{equation}\begin{split}\label{eq:infidelity_expansion}
   \mathcal{I}(T) = 1 & - | \bra{\psi_B}  U(T) \ket{\psi^0_{A+\delta}} \braket{\psi^0_{A+\delta}}{\psi_A}  \\
   & +  \sum_{i>0}  \bra{\psi_B}  U(T) \ket{\psi^i_{A+\delta}} \braket{\psi^i_{A+\delta}}{\psi_A} |^2 
  \end{split}
\end{equation} which allows for an approximate analysis of the separate contributions. Here we focus on the groundstate contribution (first line), making the assumption that the second line is small in comparison.

To approximate the groundstate contribution we insert projections onto the groundstate $\ket{\psi^0_{v}}$ of a moving frame Hamiltonian $H(v)$ with a velocity $v$ equal to the bulk velocity of the strategy and finally a projection onto the groundstate $\ket{\psi^0_{B-\delta }}$ of the Hamiltonian Eq. \ref{eq:Kitaevchain} with the wall at position $x_{L}=x_{B}-\delta$ (just before the final jump) resulting in
\begin{eqnarray}\label{eq:infidelity_groundstate_projections}
    \mathcal{I}(T) &  \approx &1- |\braket{\psi_B}{\psi^0_{B-\delta } } \braket{\psi^0_{B-\delta} }{\psi^0_v } \times \\
   && \bra{\psi^0_v } U(T)\ket{\psi^0_v} \braket{\psi^0_v}{\psi^0_{A+\delta }} \braket{\psi^0_{A+\delta }}{\psi_{A} }|^2.
 \nonumber 
 \end{eqnarray}
 The $\braket{\psi_{A(B)\pm\delta}}{\psi_{A(B)}}$ represent the initial and final jumps in position space of size $\delta$ which by fitting to our numerical model can be characterized  as \begin{equation}
O_\delta = |\braket{\psi_{x_L} }{\psi_{x_L+\delta }}|^2 \sim {\exp( - \delta^2/ s^2)}
\label{eq:Od}
\end{equation} where $s \sim \epsilon+ \alpha \lambda_F$ with $\lambda_F$ the Fermi wavelength and the fitting parameters $(\epsilon,\alpha) = (-0.33,0.44)$ for one-wall and $(-0.12,0.3)$ for two-walls when $\Delta=0.3$. In the appendix \ref{app:JinfCost} we discuss this in more details, showing also how the fit $s$ scales with the coherence length $\xi$ of the Majorana mode.

The amplitude $\braket{\psi^0_{v}}{\psi^0_{A+\delta }}$ represents the overlap between the static ground state and that of a moving frame with a constant velocity $v$. We estimate this as follows:
\begin{equation}
 O_{v} = |\braket{\psi^0_{x_L} }{\psi^0_{v}}|^2 \sim  1- \beta \left( \frac{\nu}{\Delta} \right)^2 \gamma
 \label{eq:Ov}
\end{equation}
with $\gamma = 1/\sqrt{1- v^2 /\Delta^2} $. One way to argue for this form is to use the results of \cite{Conlon2019} that showed that the energy of the moving frame ground state w.r.t. to the static frame goes as $E \sim \gamma M v^2 $ where $M \propto k_F/\Delta $. The overlap can be related to this energy via  $O_v \sim1 - E/ k_F \Delta $. We arrive at a value of $\beta \sim 0.13$ by fitting Eq. \ref{eq:Ov} to moving frame (two-wall) numerical simulations for $v \ll v_{\text{crit}}$.  For single wall motion we can calculate  value of $\beta \sim 0.065$ by slowly accelerating the wall up to  $v \ll v_{\text{crit}}$ and comparing the evolved state with the instantaneous ground-state. Note that this analysis also holds for the more general proximity coupled semiconducting nanowire with different values for fitting parameters $\alpha$ and $\beta$.  

\begin{figure}[h!]	
	\includegraphics[width=1.0\linewidth]{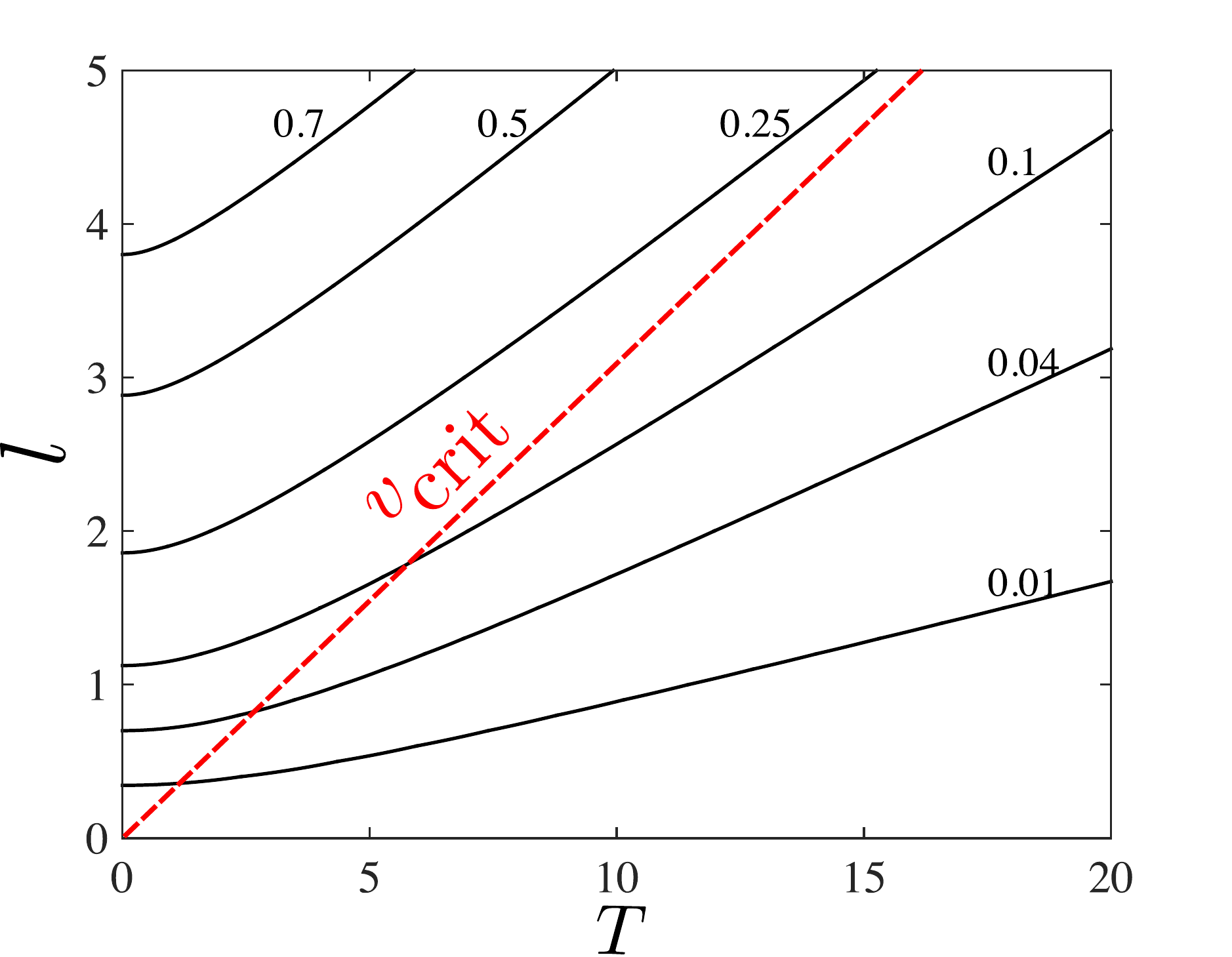}
	\caption{Contours showing the infidelity  $\mathcal{I}$ for the bare (one-wall) jump-move-jump strategy (Eq.~\ref{eq:maximizingfidelity}) as a function of distance $l$ and time $T$ using the same fitted parameters as Fig.~\ref{fig:Fexplain}. The red dashed line indicates when $v_{\text{avg}}=l/T=v_{\text{crit}}$ which can be used to distinguish the different Majorana motion regimes as defined in Fig.~\ref{fig:majorana_game_setup} (c). The jump-move-jump protocol allows for low infidelities even in cases where the average velocity exceeds $v_{\text{crit}}$ (regime I).}
		\label{fig:Isurface} 
\end{figure}

If we assume a symmetric strategy (e.g. jumps of the same size at the beginning and end) then the fidelity function $\mathcal{F}({T}) \approx |O_\delta O_\nu|^2$ (where $\mathcal{I}({T}) = 1- \mathcal{F}({T})$) to be maximized is approximated as
\begin{equation}
\label{eq:maximizingfidelity}
\mathcal{F}(T) \approx  \exp( - \frac{(l- v T)^2}{2s^2}) \times \left[1- \beta \gamma \left( \frac{v}{\Delta} \right)^2 \right]^2
\end{equation}
where we have substituted $\delta = (l -vT)/2$.  The first function is maximized by making $v$ as large as possible while the second prefers to have small $v$, with a very severe penalty kicking in as $v$ approaches the critical velocity~$\sim \Delta$. In Figure \ref{fig:Fexplain} we show how this function behaves for the protocol in regime I, which indicates that the analysis accurately captures the behaviour of the fidelity at small $v/\Delta$ but also displays overall qualitative agreement with the numerical results for a wide range of values of $v/\Delta$ including the presence of a maximum in the fidelity.

An important part of the strategy are the jumps at the beginning and end of the protocol that allow for a constant middle motion at lower velocities. In Fig.~\ref{fig:Isurface} we show how the infidelity behaves for a range of movement lengths $l$ and times $T$ obtained from maximizing Eq. \ref{eq:maximizingfidelity} with respect to $v$. By comparing with Fig. \ref{fig:majorana_game_setup} (c) we see that in regime I the bare JMJ strategy gives infidelities of $\mathcal{I}(T)>0.25$, in regime II (between the dashed lines) about $0.04<\mathcal{I}(T)<0.25$ and in the short $l$- short $T$ regime III (below the blue dashed line) performs particularly well with infidelities $\mathcal{I}(T)\sim\mathcal{O}(10^{-2})$. 

This method is only effective however if the jumps needed are not too large ($\delta \sim O(\lambda_F)$).  Interestingly, as the overlap behaviour of $O_v$ is relatively unaffected by changes to the chemical potential, this suggests that a chemical potential nearer to the bottom of the band, $\mu \rightarrow -2 w$ can help to extend the range of the protocol.

\subsection{Dressed Jump-Move-Jump ($\Delta x_{\text{back}} \neq 0$)}
\label{sect:xbn0}
\begin{figure}[h!]	
	\includegraphics[width=1.0\linewidth]{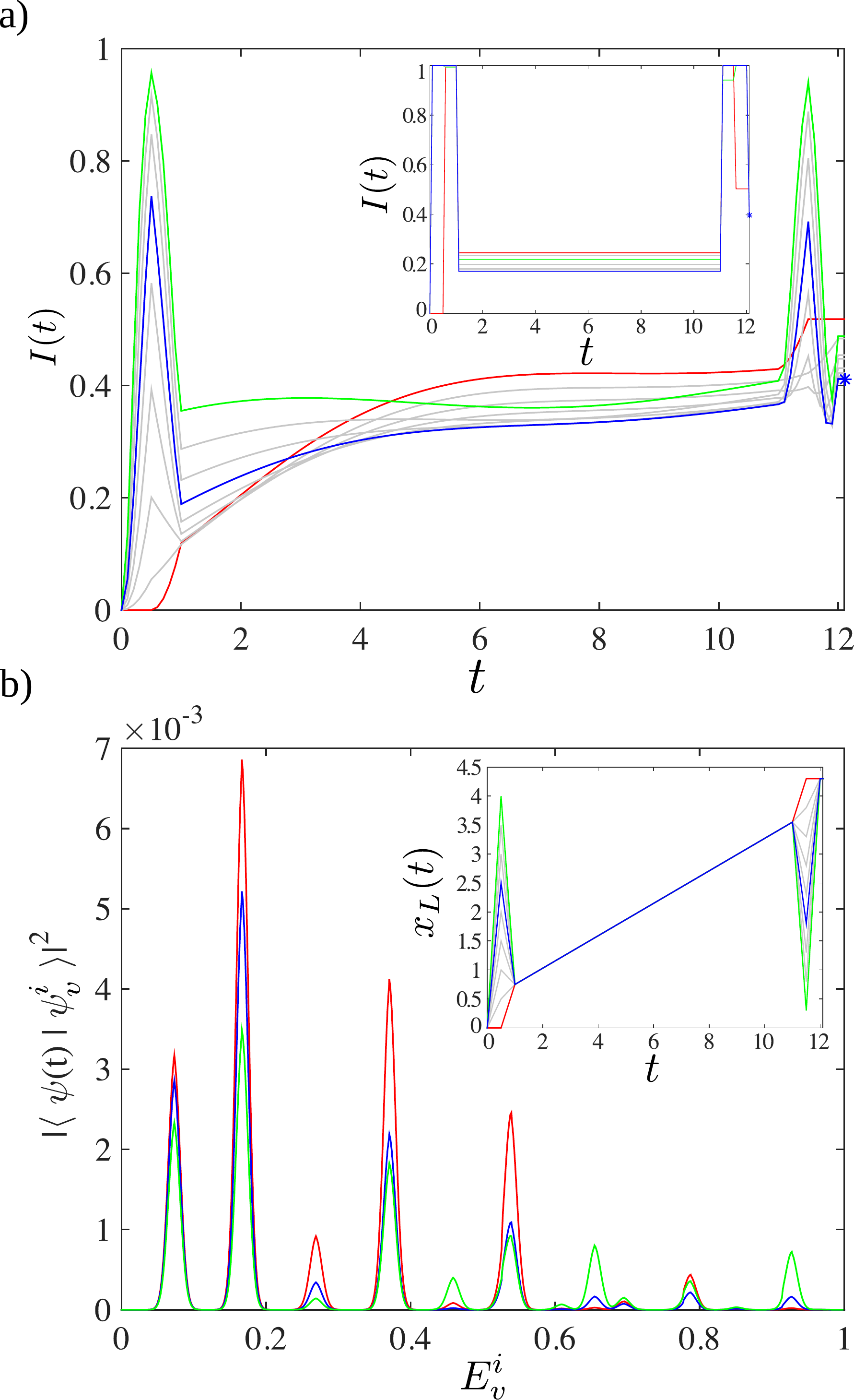}
	\caption{a) Instantaneous Infidelity $I(t)$ with respect to the static (lab) frame groundstate (main panel) and moving frame groundstate (inset) as a function of the protocol time $t$ for various dressed jumps in regime I which are depicted by the different colours (red,blue,green,grey) and shown in the inset of b). The optimal dressed jump (blue) has the minimum infidelity with the moving frame groundstate and becomes the strategy with the minimum static frame infidelity after about $t\approx4$. b) The time-evolved state $\ket{\psi(t)}$ expanded in the single and two particle excitations $\ket{\psi^i_v}$ of the moving frame basis with energies $E^i_v$ at $t=T/2$ for a forward jump (red), an optimal dressed jump (blue) and non-optimal dressed jump (green) for regime I as given in the inset. The occupation probabilities (delta functions) are convoluted for visualisation purposes. The forward jump strategy (red) occupies the excitations close to the bottom of the band in the moving frame $(E^i_v\approx0.08)$ more compared to the dressed jumps (blue and green). In this figure all times are in units of $1/\omega$.}
		\label{fig:dressed_jumps} 
\end{figure}

On top of the bare jump-move-jump structure, the ML algorithms dress the protocols with additional forward and backward movements. In this section we will show that the primary purpose behind these additional motions is to better target the moving frame ground state in the \textit{move} part of the \textit{jump-move-jump} protocol.

To proceed we focus on the simple upgraded JMJ protocol that also allows backward motions over distances $\Delta x_{\text{back}}\neq0$, see Fig.~\ref{fig:simple_model} (a). Our first key observation  is that the optimal protocols here tend to choose a combined jump size $\Delta x_{\text{forward}} - \Delta x_{\text{back}} \sim C$, see Fig.~\ref{fig:simple_model} (b), where $C$ is a fixed constant. This indicates the same primary goal as the JMJ strategy: to allow a period of optimal and roughly constant sub-critical motion.

To understand why the backward motion is a further benefit we examine the 
instantaneous infidelity in both the static and moving frames for a series of protocols with the fixed bulk velocity $v$, see Fig.~\ref{fig:dressed_jumps} (a). The instantaneous infidelity is defined as $I(t) = 1 - |\braket{\psi(t)}{\psi_{\text{ins}}^0(t)}|^2$ with $\ket{\psi(t)}=\mathcal{T}e^{-i\int_0^tH(t')dt'}\ket{\psi_A}$ in which $H(t)$ follows the time-dependence of the dressed JMJ strategy and $\ket{\psi_{\text{ins}}^0(t)}$ corresponds to the instantaneous groundstate (at the fixed time $t$) of the static frame Hamiltonian $\mathcal{H}(t)$ (Eq. \ref{eq:Kitaevchain}) or the moving frame Hamiltonian $H_v(t)$ (Eq. \ref{eq:movHam}). The static frame perspective shows that the optimal jump finds a state that slowly becomes the lowest infidelity state after some time. However a more revealing picture emerges if we examine the same set of protocols in a moving frame (inset). Here we see that the objective of the initial jumps is to maximize the overlap with the ground state of the moving frame. 

Using the same methodology we can also reveal what moving frame bulk excitations $\ket{\psi^i_v}$ with many-body energy $E^i_v$ (restricting here to single- and two particle excitations only) the dressed protocol occupies during the evolution, see Fig.~\ref{fig:dressed_jumps} (b). For protocols close to the bare JMJ strategy (shown in red) one finds that the dominant excitations have energies close to the moving frame bulk gap $(E^i_v\approx0.08)$. The dressed jumps (in blue and green) however are able to lower the amplitudes of these excitations, but at the cost of also exciting some higher energy quasi-particles. For dressed jumps that are too large (in green) these higher energy excitations eventually dominate and one gets an increase in the ground state infidelity.  

Before ending this section, we now briefly summarize why similar JMJ strategies work in the regimes I-III. The first advantage of the JMJ strategy is that, with abrupt initial and final costs, it allows for an effective reduction in the distance $l$. This is especially obvious in the case of regime I where, by definition, simple linear motion implies a need to move above the critical velocity.  

The other crucial aspect of the dressed protocols is that they can better target the ground state of the moving system through the combination of jumping and moving.  This allows the system to be rapidly accelerated to a near constant velocity frame where there is no additional infidelity cost. For a fixed distance $l$, the JMJ strategy therefore generally allows for a reduced time by cutting out the slow acceleration parts of regime IV, allowing regime III, regime II and, to a lesser extent, regime I exploit the property of super-adiabaticity.  

Of course it is always possible to define scenarios deep in regime I where the distance and time constraints would lead to large infidelity even for the dressed JMJ protocols.   However they do allow, under the same spatial and temporal constraints, for Majoranas to be moved more efficiently than that of linear motion or slow ramp-up ramp-down protocols.
 
\section{Robustness of the JMJ protocol with respect to interactions and disorder} \label{sec:robustness} 

In the space between our simplified theoretical setups (the Kitaev chain in Eq. \ref{eq:Kitaevchain} and Proximity Coupled Semiconductor in App. \ref{app:semiconductor}) and actual topological quantum devices there are many additional layers of complexity, and one might wonder which aspects of the JMJ protocol remain robust. In this section we assess the robustness of our optimal JMJ protocols obtained in the clean non-interacting system when applied in a system with interactions (see e.g \cite{Stoudenmire2011, Sela2011, Lutchyn2011,Crepin2014,Gergs2016,Kells2018,Moon2018}), and separately applied in a system with disorder, see \cite{Brouwer2011,Brouwer2011b,DeGottardi2013}. We find that the fidelity loss due to the jumps of the JMJ protocol remains robust and that the overall protocol still outperforms the naive linear benchmark protocol significantly. Finally we comment on the extension of our ML methods to include these more complex scenarios. 
\begin{figure}[h!]	
	\includegraphics[width=1.0\linewidth]{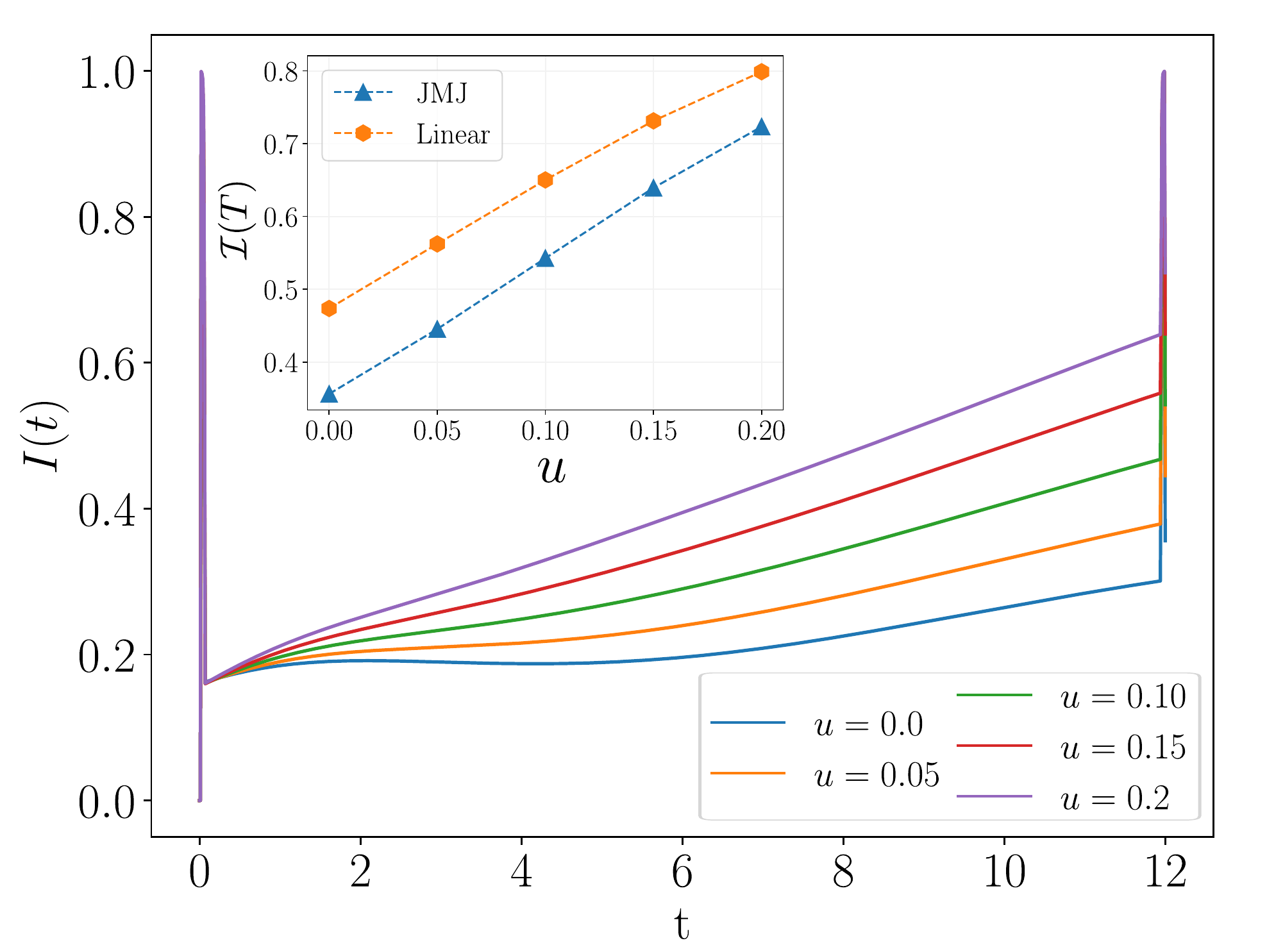}
	\caption{Main panel: instantaneous infidelity $I(t)$ of the optimal JMJ protocol from the non-interacting system in regime I run in a system with several different finite interaction strengths $u$. The jumps are robust against interactions whereas infidelity increases during the move part of the protocol. Inset: the final infidelity values $\mathcal{I}(T)$ increase approximately linearly with interaction strength for both a linear as well as the best JMJ protocol. For all interaction strengths the JMJ protocol outperforms the linear protocol. The other parameters for these simulations are the same as in Fig. \ref{fig:overview} and the parameters of the JMJ protocol can be found in App.~\ref{app:model}.}
		\label{fig:int_reg1} 
\end{figure}

\subsection{The effect of interactions}

Interactions are modeled by adding the nearest neighbour density-density interaction term $ H_{\text{int}}=\sum_{x=1}^{N-1} u_x c^\dagger_x c_x c^\dagger_{x+1} c_{x+1}$ with interaction strength $u_{x}$ to the Kitaev chain Hamiltonian in Eq. \ref{eq:Kitaevchain}, $H(t)=\mathcal{H}(t)+H_{\text{int}}$. Since $H_{\text{int}}$ is quartic in the creation and annihilation operators, the BdG time evolution algorithm used before (see App. \ref{app:gradient_analysis}) cannot be applied directly and we consider a time-dependent variational principle with  matrix product states (MPS-TDVP) \cite{HaegemanTDVP, Paeckel2019} approach to simulate the dynamics. 

In Fig. \ref{fig:int_reg1} we show how the optimal protocol obtained in the non-interacting system in regime I performs for various finite interaction strengths (see App. \ref{app:int_dis} for the other regimes). From the instantaneous infidelity plot in the main panel we can see that the fidelity loss due to the jumps remains robust with increasing interaction strength whereas in the bulk (move part) the loss increases. As a consequence the final infidelity $\mathcal{I}(T)$ increases with interaction strength (See inset Fig.~\ref{fig:int_reg1}), but the performance remains superior to a naive linear protocol at all interaction strengths. The increase of infidelity with interaction can be argued for on the basis of mean-field theory, where one expects repulsive interactions to lower the effective gap (leading to a lower critical velocity) and the appearance of non-uniformity in the effective coupling parameters near the system boundary.  

Since we did not perform the optimization directly in the presence of interactions, we cannot argue that the optimal JMJ protocol from the non-interacting system is also the optimal protocol in the interacting system. We note however that our ML optimization methods can be directly extended for this task. Firstly, the NES algorithm is a black-box optimization method and can be combined with any method that generates the dynamics such as the MPS-TDVP algorithm. Secondly, to apply DP one needs to make the code for the MPS-TDVP algorithm differentiable. In this regard we note that the DMRG algorithm to find ground states in interacting systems has been recently made differentiable~\cite{Liao2019}. 

\begin{figure}[h!]	
	\includegraphics[width=1.0\linewidth]{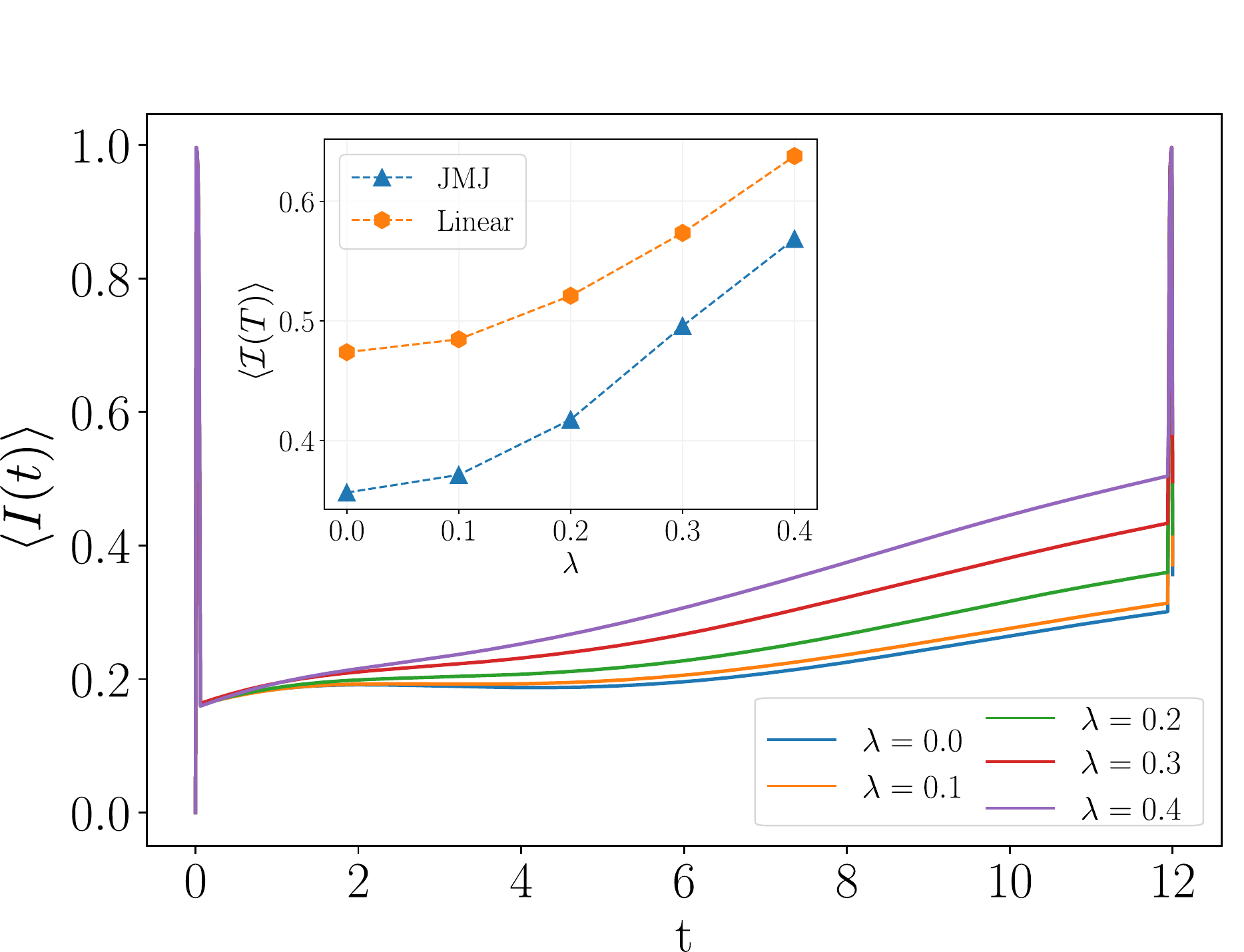}
	\caption{Main panel: disorder averaged instantaneous infidelity $\langle I(t)\rangle$ of the optimal JMJ protocol from the clean wire in regime I for different disorder strengths $\lambda$. The jumps are robust against weak disorder whereas the infidelity increases during the move part of the protocol. The disorder averaging was done over 500 realizations. Inset: the final infidelity values $\langle \mathcal{I}(T)\rangle$ increase approximately quadratically with disorder strength for both a linear as well as the best JMJ protocol. For all disorder strengths the JMJ protocol outperforms the linear protocol. The other parameters for these simulations are the same as in Fig. \ref{fig:overview} and the parameters of the JMJ protocol can be found in App.~\ref{app:model}.}
		\label{fig:dis_reg1} 
\end{figure}
\subsection{The effect of disorder}\label{effectdisorder}

Disorder in the wires can be modeled by adding a random Gaussian noise term $\Tilde{\mu}(x)$ with mean $0$ and standard deviation $\lambda$ to the chemical potential $\mu_{\text{dis}}(x)=\mu(x) + \Tilde{\mu}(x)$ \cite{Brouwer2011,Brouwer2011b,DeGottardi2013}. It has been shown that this form of disorder leads to an increase in the effective coherence length $\xi_{\text{eff}}$ of the Majorana, which in the continuum limit leads to $\frac{1}{\xi_{\text{eff}}} = \frac{1}{\xi} - \frac{1}{2l_{\text{dis}}}$. Here $l_{\text{dis}}=v_{\text{F}}^2/\lambda^2$ is the characteristic disorder length scale. When the disorder strength is such that $2l_{\text{dis}}=\xi$, the effective Majorana coherence length $\xi_{\text{eff}}$ diverges and the topological phase is destroyed. Another consequence of disorder is that the critical velocity is significantly reduced~\cite{Karzig2013} and the ground state fidelity is expected to rapidly decrease upon smoothly moving the Majoranas over a disordered medium.

In Fig. \ref{fig:dis_reg1} we show the performance of the JMJ protocol in regime I for various different disorder strengths  compared to a linear benchmark protocol. Regimes II, III and IV are detailed in App.~\ref{app:int_dis}. From the averaged instantaneous infidelity $\langle I (t) \rangle$ in Fig. \ref{fig:dis_reg1} it can be seen that the infidelity associated with the jumps is robust to the presence of disorder but there is a relatively strong fidelity loss associated with the linear motion of the move part. This is in line with the expectations given the lower critical velocity and longer coherence length induced by disorder. While the final infidelity value $\mathcal{I}(T)$ of the JMJ protocol remains reasonably high in regime I, for longer time/distance regimes (e.g. deep in regimes II/IV) $\mathcal{I}(T)$ worsens significantly, which suggests that additional protocol optimization in the presence of disorder may be required to achieve comparable infidelities.  

The protocol optimization in the presence of disorder can be achieved through the NES method since the computational complexity grows linearly with the number of disorder realizations, all of which can be simulated in parallel. The DP method can also be applied because the disorder averaging is a differentiable operation, i.e., DP can access  the disorder averaged infidelity $\nabla_{\theta}\langle \mathcal{I}(T)\rangle$ with respect to the control parameters. For example \cite{schaefer2021control} and \cite{PhysRevA.99.052327} consider DP for stochastic optimization tasks.

\section{Conclusion and further work}

We have applied two state-of-the-art machine learning techniques, namely differentiable programming (DP) and natural evolution strategies (NES), to the problem of manipulating Majorana bound-states in a topological superconductor. For DP we have shown that the entire dynamical evolution of any free-fermion system can be functionally differentiated, allowing for efficient optimization protocols. This in turn provides an ability to tackle computationally harder problems and allows the dynamical optimization to be integrated seamlessly with both direct and neural-network parameterizations of quantum control protocols. 

In addition to this we have shown how the Majorana control problem itself can be naturally formulated as a game. This allows the application of reinforcement learning approaches and NES to zero-mode manipulation.  The key advantage here is both speed and flexibility compared to, e.g., the standard MC methods such as simulated annealing.
Beyond the advantages observed in our numerical experiments there are a number of additional conceptual and practical ingredients which can be taken advantage of in the reinforcement learning setup including a flexible exploration step, delayed rewards, partial observations, and the ease of accounting for the stochastic nature of the reward function, all of which may be particularly relevant to experimental setups.

Note that these machine learning approaches are not specific to Majorana bound state control, and could also be applied to different dynamical many-body problems with radically different motivations and cost functions (e.g. log-likelihood~\cite{Bishop2006}, Fisher information~\cite{petz2011} or the trace distance~\cite{Nielsen}). 

On the theoretical level we have introduced a framework for dividing up  the Majorana control problem into four distinct transport regimes. This has deepened the understanding of the problem and help to establish a connection with optimal control theory more broadly. Remarkably our numerical simulations displayed an innate awareness of this rich landscape and uncovered important hidden aspects of the underlying physics. In particular they have identified a new class of protocol that radically outperforms other known strategies (adiabatic and bang-bang) when there are both spatial and temporal constraints. We have shown that this family of protocols, and the associated classification of Majorana motion, also holds for the richer proximity-coupled semiconductor nanowire model, which is directly relevant for experiments. In principle, since the defining criteria are related to topological gap, similar protocols should also apply to related models such as the SSH chain \cite{PhysRevLett.42.1698} or multi-channel topological wires variants  \cite{Rieder2012}.  

The core of this alternative strategy is a sudden jump followed by a period of constant velocity motion and then another jump. A theoretical analysis of this dynamics shows that the strategy simultaneously exploits the models stability at constant velocities below a certain threshold, and the property that small instantaneous changes in the domain wall position do not dramatically reduce the ground state fidelity. Further analysis of dressed JMJ protocols reveals that more complex initial jump sequences can, in addition to allowing a sub-critical velocity, better target the ground state of the moving frame. 

The protocol resembles the parallel recent development of the  Bang-Anneal-Bang protocols \cite{Brady2020} for ground state preparation in Noisy Intermediate Scale Quantum devices. Our theoretical analysis illuminates why this type of strategy works so successfully and we are hopeful that this approach will motivate more general theoretical analysis of related protocols. In this context it would also be interesting to see the same types of protocols also emerge in more realistic materials. Our results for the semiconductor model suggest that similar protocols are indeed relevant. However, in that example, it is important to note that boundary motion does not strongly couple between bands with different spin orientations. In real materials one might expect that sudden boundary change would couple the ground state to bands at larger energies and this may restrict the range of motions that one can achieve.

Another natural direction to explore is the application of our methods to the problem of Majorana braiding, which is the ultimate goal behind the study and control of zero-modes in the context of topological quantum computing. It remains an interesting open question whether aspects of our protocols remain robust if one models specific devices more closely.  We emphasize however that the strength of the RL setup is that one can easily encode other experimental restrictions on the control by including them in the reward or action spaces \cite{6918520}. As such the application of these methods can be readily extended to other more complicated models, such as models including the effect of interactions~\cite{PhysRevB.84.014503}, different simulation techniques such as  the time-dependent density-matrix renormalization group~\cite{PhysRevLett.93.076401}, and to measurement-based Majorana devices~\cite{Karzig2017}.

In a broader sense we believe our work shows that machine learning for quantum control can be done at a scale, speed, and precision that is relevant to modern experimental devices.  As such we believe that it will be used to overcome obstacles to realizing large-scale quantum computation, quantum communication networks, quantum thermal machines, analogue quantum simulations, and control of other quantum many-body dynamical systems more broadly.

\begin{acknowledgements}
We would like to thank Andreas Buchleitner, Steve Campbell, Aaron Conlon, Shane Dooley, Ian Jubb, Kevin Kavanagh, Jinguo Liu, Xiuzhe Luo, Frank Sch\"{a}fer, Jiayu Shen and Lei Wang for helpful discussions regarding this project.  L.C and G.K. acknowledge Science Foundation Ireland for financial support through Career Development Award 15/CDA/3240. G. K. was also supported by a Schr\"{o}dinger Fellowship. J.C. acknowledges support from Natural Sciences and Engineering Research Council of Canada (NSERC),  the Shared Hierarchical Academic Research Computing Network (SHARCNET), Compute Canada, Google Quantum Research Award, and the Canadian Institute for Advanced Research (CIFAR) AI chair program. D.L and BKC acknowledge support from the Department of Energy grant DOE desc0020165.
\end{acknowledgements}

\bibliography{reference}

\clearpage

\onecolumngrid
\begin{center}
	\noindent\textbf{Appendix}
	\bigskip
		
	\noindent\textbf{\large{Protocol Discovery for the Quantum Control of Majoranas by Differentiable Programming and Natural Evolution Strategies}}
		
\end{center}

\appendix

\renewcommand\thefigure{A\arabic{figure}}  
\renewcommand\thetable{A\arabic{table}}  
\setcounter{figure}{0}  
\setcounter{table}{0}

\section{\label{app:majorana_motion} Derivation of the Critical Velocity and Moving Frame Picture}

In this appendix we first introduce some background theory related to the critical velocity and the Majorana-mode wave function in the moving frame. Afterwards we derive the resonance time scale $T_{\text{res}}$ for Majorana shuttling from the Fermi golden rule and finally we show the emergence of this resonance timescale in numerical simulations for forward oscillating motion. 

\subsection{Dispersion and Majorana Mode in the Moving Frame}
To find the critical velocity we start from the periodic (N+1=1) Kitaev chain Hamiltonian (Eq. \ref{eq:Kitaevchain} with $V(x,t)=0$) in the continuum which reads after a Fourier transformation \begin{equation}\label{eq:Hamcont}
    H_k = \Vec{d}_k\cdot\Vec{\sigma} = \begin{pmatrix}
\frac{k^2}{2m}-\mu & i\Delta k  \\
-i\Delta k & -\frac{k^2}{2m} + \mu  
\end{pmatrix}
\end{equation} with $\Vec{d}_k=(0,-\Delta k, \frac{k^2}{2m}-\mu)$, momentum $k$  and $\Vec{\sigma}$ the Pauli spin matrices. Solving for the energy eigenvalues gives mode dispersion $\epsilon_k = \pm\abs{\Vec{d}_k}=\pm\sqrt{(\frac{k^2}{2m}-\mu)^2+\Delta^2k^2}$. 
The moving frame can now be obtained by a Galilean transformation in terms of the unitary rotation $U(t)=e^{-ik\int_0^t v(t')dt'}$ which is the time dependent translation operator that rotates to a frame with domain wall velocity $v(t)$. Applying this to \ref{eq:Hamcont} gives the moving frame Hamiltonian 

\begin{align}
     H_v(t) = U^\dagger(t) H_k U(t) + i \frac{dU^\dagger}{dt}U(t) =  \begin{pmatrix}
 \frac{k^2}{2m}-\mu + v(t)k & i\Delta k  \\
 -i\Delta k & -\frac{k^2}{2m} + \mu +v(t)k 
 \end{pmatrix}.
  \label{eq:movHam}
 \end{align}

\begin{figure}[h!]
    \centering
    \includegraphics[width=1.0\linewidth]{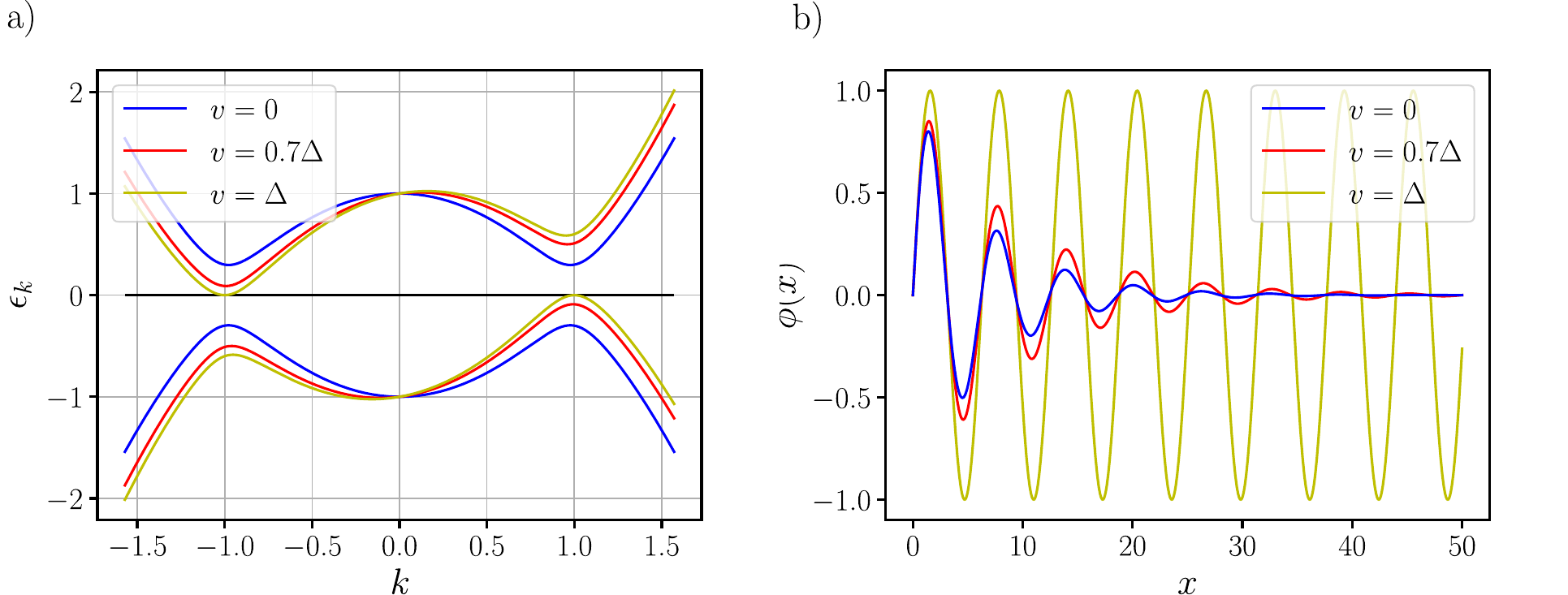}
    \caption{a) Mode dispersion in the moving frame for several different velocities. b) Localization of the Majorana zero modes in the moving frame for several different velocities. It can be seen that the gap closes for $v=v_{crit}=\Delta$ and simultaneously the Majoranas delocalize in the moving frame.}
    \label{fig:dispersionMajorana}
\end{figure}

The moving frame has an extra term $v(t)k$ on the diagonal which means the moving frame energy dispersion is $\epsilon_k = \pm\abs{\Vec{d}_k}=\pm\sqrt{(\frac{k^2}{2m}-\mu)^2+\Delta^2k^2}+vk$ ~\cite{Scheurer2013}. This results in a tilted dispersion as shown in Fig. \ref{fig:dispersionMajorana},  when $v=\Delta$ the tilt causes the gap to close and we have a topological phase transition. This velocity defines the critical velocity $v_{\text{crit}}=\Delta$.

Besides the energy dispersion, we can also look at the wave function of the Majorana zero modes in the moving frame. To find this wave function we solve for the zero-energy solutions of the moving frame Hamiltonian \ref{eq:movHam} which results in $\psi=[\phi,-\phi]^\intercal$ with 
\begin{equation}
    \phi(x) \propto  e^{- x/(\gamma\xi)}\sin(\sqrt{k_F^2 +1/(\gamma\xi)^2}x).
\end{equation} The difference with the Majoranas in the lab frame is that the localization length $\xi=1/\Delta m$ is dilated by a factor $\gamma=1/\sqrt{1-\frac{v^2}{\Delta^2}}$ causing the Majorana modes to become spatially more extended (delocalized) in the moving frame. When $v=v_{\text{crit}}=\Delta$ the localization length becomes infinite indicating that the local character is lost and hence a topological phase transition occurs as shown in Fig. \ref{fig:dispersionMajorana}.  

\subsection{Derivation and numerical analysis of the resonant time scale}

To derive the time scale for resonant Majorana motion we consider a scenario in which we are shuttling the left Majorana back and forth by using \begin{equation}
    v_L(t) = v_{\text{max}}\sin{\omega t}
\end{equation} for the velocity of the left domain wall in Eq. \ref{eq:Potentialprofile}. When $v_{\text{max}}$ is not too large, i.e. the amplitude of the left domain wall position can be considered small compared to the total length of the wire $N$, we can treat this motion as a time-dependent perturbation and apply perturbation theory. In this case we write the external potential as $V=V_0+\delta V \sin{\omega t}$ in which $V_0$ is the static domain wall profile in Eq. \ref{eq:Potentialprofile} and $\delta V \sin{\omega t}$ a time-dependent fluctuation on top of it. From Fermi's Golden Rule for a harmonic perturbation~\cite{Sakurai:1167961} we can find the infidelity rate to be \begin{align}
    \lim_{T\mapsto\infty}\frac{\mathcal{I}(T)}{T} = \lim_{T\mapsto\infty}\frac{1}{T}[1 - |\bra{\psi_0}\exp{-i\int_0^T \mathcal{H}(t)dt}\ket{\psi_0}|^2] = \lim_{T\mapsto\infty}\frac{1}{T}\sum_{i\neq0} |\bra{\psi_i}\exp{-i\int_0^T \mathcal{H}(t)dt}\ket{\psi_0}|^2 = \\ 2\pi\sum_{i\neq0}(\delta(E_i-E_0+\omega)+\delta(E_i-E_0-\omega))|\bra{\psi_i}\delta V\ket{\psi_0}|^2
\end{align} in which $\ket{\psi_i}$ are the eigenstates of the Hamiltonian with the static domain wall profile $V_0$ which have energy $E_i$. 

From this equation we can read off that there are resonances in the infidelity whenever $\omega_{res}=\pm(E_i-E_0)$, for the first excited state $\omega_{res}=E_i-E_0=\Delta k_F$ is equal to the gap in the system and we arrive at the resonance time $T_{res}=\frac{2\pi}{\omega_{res}}=\frac{2\pi}{\Delta k_F}$. The resonances for the higher energy bands are suppressed with the transition amplitude $|\bra{\psi_i}\delta V\ket{\psi_0}|^2$. When $\omega\mapsto \infty$ there are no resonances and the infidelity becomes zero; the intuition for this is that the static system $H_0$ does not notice the harmonic perturbation because it oscillates at a much higher frequency. Similarly, when $\omega<\omega_{res}$ there are also no resonances because the frequencies are smaller than the gap which corresponds to adiabatic Majorana motion.  

\begin{figure}[h!]
    \centering
    \includegraphics[width=1.0\linewidth]{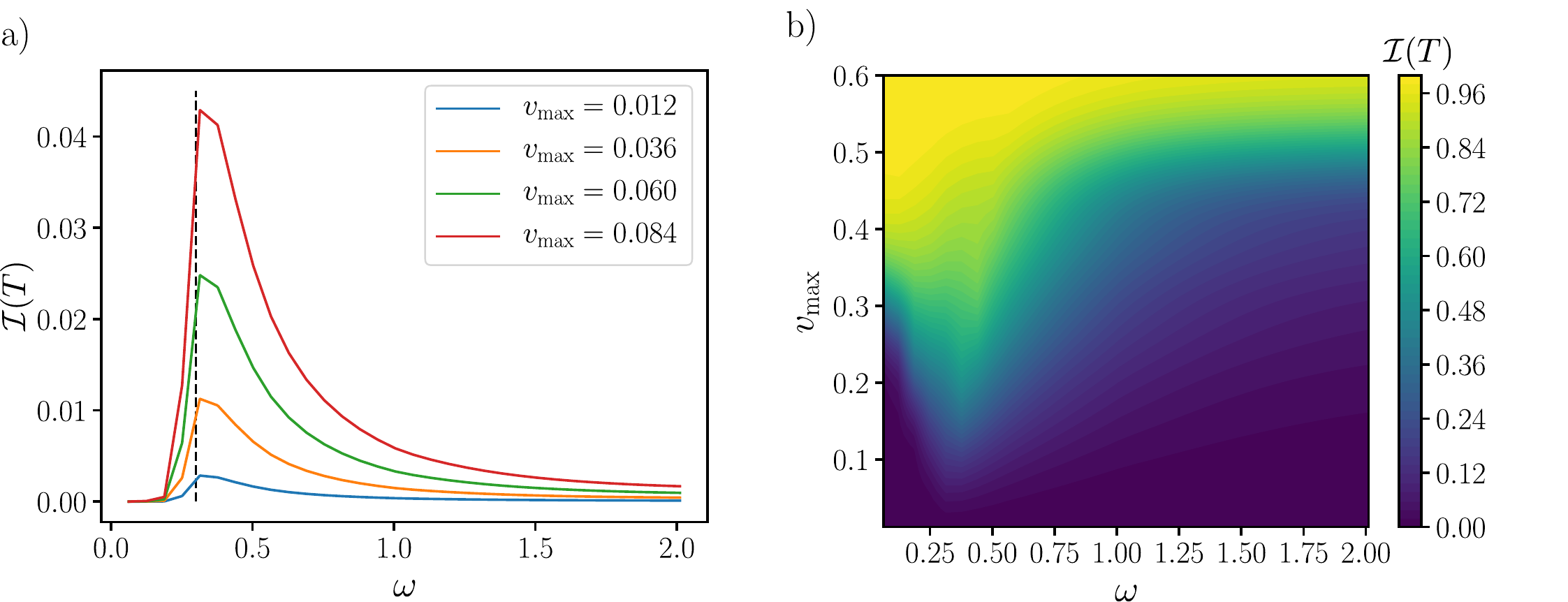}
    \caption{a) Infidelity as a function of frequency $\omega$, the black dashed line indicates the resonance frequency $\omega_{\text{res}}=\Delta k_F=0.3$. At $\omega\mapsto\infty$ the infidelity is the same as for constant linear motion with $x_L(t)=\frac{v_{\text{max}}}{2}t$. b) Infidelity as a function of frequency and $v_{max}$. For velocities $v_{\text{max}}>v_{\text{crit}}=0.3$ there are no low adiabatic frequencies anymore. The parameters used in these simulations are $N=140$, $\mu=1$, $w=1$, $\Delta=0.3$ and $\sigma=1$.}
    \label{fig:resonance}
\end{figure}

In the Majorana motion problem in the main text we are however looking at moving the left domain wall forward from $x_A$ to $x_B$. As shown in Fig. \ref{fig:resonance} we see that the same resonance time scale emerges for forward motion with \begin{equation}
    v_L(t) = v_{\text{max}}\frac{1-\cos{\omega t}}{2}.
\end{equation} For $\omega\mapsto\infty$ we get the same rate as constant motion with $\frac{v_{\text{max}}}{2}$ which can be explained for low velocities from adiabaticity. 

\section{Algorithm and Analysis of the Derivative of the Infidelity}\label{app:gradient_analysis}

In this appendix we first outline the algorithm we use to compute the many-body overlap Eq. \ref{eq:infidelity} for the Majorana wire setup. Afterwards we argue that all operations in this algorithm are differentiable and we can obtain the derivative of the fidelity with respect to control parameters using automatic differentiation. In the end we show the numerical obtained derivatives and check them with finite difference methods. 

To find the fidelity we first write the Hamiltonian Eq. \ref{eq:Kitaevchain} in Bogolyubov-de-Gennes (single-particle) form \cite{Bernevig2013,Ring2004} \begin{equation}
   \mathcal{H}(t) =  \frac{1}{2} C^\dagger H_{BdG}(t)C 
\end{equation} with $C^\dagger \equiv[c^\dagger_1 ... c^\dagger_i ....c^\dagger_N \quad c_1 ... c_i ....c_N ]$ and diagonalize it \begin{equation}\label{eq:bdg_diag}
    H_{\text{BdG}}(t)W(t) = E(t)W(t)
\end{equation} in which the columns of $W(t)=\left[\begin{array}{cc} u(t) & v(t)^* \\ v(t) & u(t)^* \end{array}  \right] $ are the eigenmodes (quasi-particle modes) of the BdG Hamiltonian and $E(t)$ is the diagonal matrix of the mode energies $[E]_{ii}(t)=\epsilon_i(t)$. 
To compute the final many-body overlap at time $t=T$  we then use the Onishi Formula \begin{equation}\label{eq:onishi}
    \mathcal{F}(T) = |\braket{\psi_B}{\psi(T)}|^2 = det(u_B^*u(T)+v_B^*v(T))
\end{equation} which relates the BdG quasi-particle picture to the many-body picture. In this $u(T)$ and $v(T)$ can be obtained from $W(T)=\mathcal{T}e^{-i\int_0^T H_{bdg}(t')dt'}W(0)$ and $u_B$ and $v_B$ from the diagonalization of the instantaneous BdG Hamiltonian with the domain wall at position $x_B$. $\mathcal{F}(T)$ can now be used to compute the infidelity $\mathcal{I}(T)$ in Eq. \ref{eq:infidelity} in the main text. 

From a programming perspective, a code to compute the infidelity from Eq. \ref{eq:onishi} and Eq. \ref{eq:infidelity}  consists of a series of elementary operations (primitives) $f_i$ that map the input control $x_L(t)$ with $t\in[0,T]$ to the output $\mathcal{I}(T)=f_n \circ f_{n-1} \circ \cdots \circ f_1(x_L(t))$. The required primitives for this algorithm are matrix multiplications, diagonalization of a hermitian matrix \ref{eq:bdg_diag} and taking the determinant \ref{eq:onishi}. Each of these operations has a known forward and reverse mode derivative \cite{Giles2008} that can be recursively assembled in the chain rule \begin{equation}
    \frac{d \mathcal{I}}{d x_L} = \frac{d f_n}{d f_{n-1}}\frac{d f_{n-1}}{d f_{n-2}}\cdots\frac{d f_1}{d x_L}
\end{equation} to find the derivative of the infidelity with respect to the control. 

\begin{figure}[h!]
    \centering
    \includegraphics[width=1.0\linewidth]{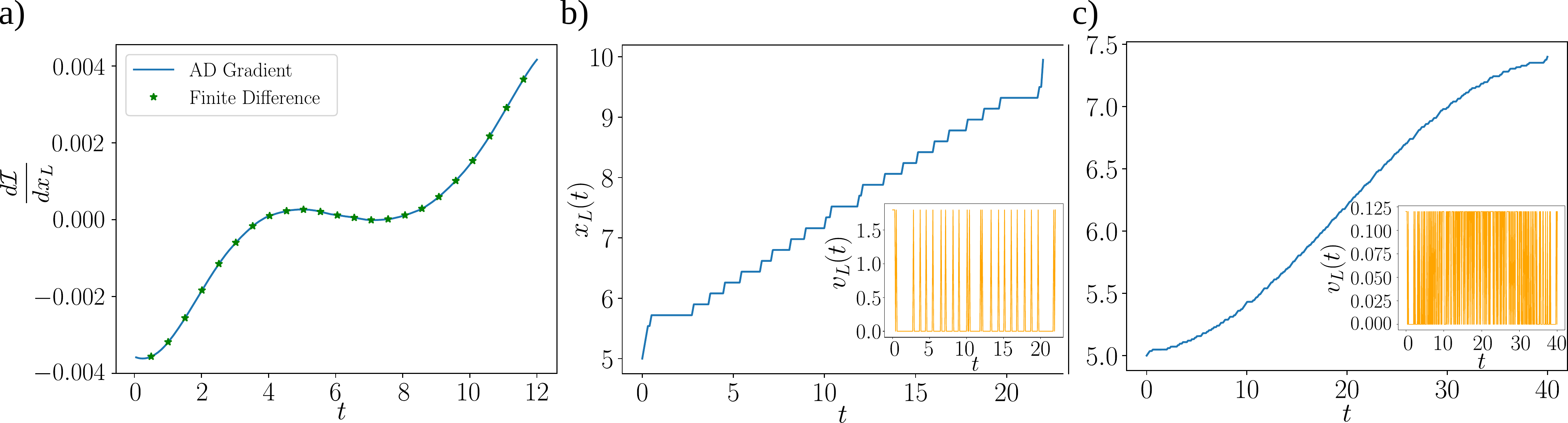}
    \caption{a) Comparison between the gradient of the infidelity $\mathcal{I}$ with respect to the control $x_L$ obtained with automatic differentiation (AD) and finite difference. The gradients $\frac{d\mathcal{I}}{d x_L}$ were evaluated for a linear protocol $x_L(t)=v_{\text{avg}}t$ as a function of time $t$ in regime I. The AD gradient matches up with the finite difference gradient and we note that at the beginning and end of the protocol the magnitude of the gradient is the largest. b) and c) Optimized parameterized bang-bang protocols in regimes II (b) and IV (c) with the simulated annealing method. The position profiles $x_L(t)$ are plotted in the main panels and the velocity profiles $v_L(t)$ in the insets. The bang-bang protocols seem to approximate the JMJ protocol in regime II and super-adiabatic protocol in regime IV. The parameters for these simulations were the same as in Fig. \ref{fig:overview} in the main text.}
    \label{fig:gradient_and_bang}
\end{figure}

An automatic differentiation package, like the JAX library \cite{jax2018github}, computes all the derivatives of the primitives in a code automatically and assembles them in the chain rule to evaluate the total gradient of a computer program. We applied this method to the code to compute the infidelity and used the gradient for the optimizations described in the main text. In Fig. \ref{fig:gradient_and_bang} a) we show an example for the gradient obtained in this way for a linear motion protocol in regime I which we checked with finite difference to make sure it was working correctly. We observe that for this protocol the gradient is the biggest at the boundary which might explain the jumping behaviour at the beginning and end of the protocols we found with the ML optimizations (Fig. \ref{fig:overview}). 

To compute the matrix exponential required for the computation of $W(T)=\mathcal{T}e^{-i\int_0^T H_{bdg}(t')dt'}W(0)$ we discretize the continuous time variable $t$ in individual time steps of size $dt$ and apply the Trotter-Suzuki expansion to write $\mathcal{T}e^{-i\int_0^T H_{bdg}(t')dt'} \approx \Pi_{j=1}^{T/dt}e^{-iH_{\text{BdG}}(jdt)dt}$. This means that while the time complexity of the back-propagation algorithm is similar to the forward evaluation, the memory complexity grows linearly with the number of discrete time steps $T/dt$. A method to circumvent this growth in memory complexity is the use of adjoint sensitivity methods \cite{chen2019neural} which avoid back propagating through the time evolution differential equation (Schr\"{o}dinger equation) by solving a second differential equation backwards in time. In practise the memory complexity did not turn out to be a bottleneck in our simulations however and it was sufficient to use the backpropagation algorithm as described above. For extending our methods to include disorder and interactions or for simulating very long times $T\sim \mathcal{O}(10^3)$ the adjoint method might be a good option to explore and we refer the interested reader to \cite{2018arXiv181201892R} for comparative studies between the adjoint and backprogagation algorithms.

\section{Gradient Formula for Natural Evolution Strategies}
\label{app:NES}
Denote $p_{\theta}(\phi)$ for the Gaussian distribution $\phi \sim \mathcal{N}(\theta, \sigma^2 I)$ with $\sigma$ fixed. It follows that \cite{JMLR:v15:wierstra14a} 
\begin{align}
     \nabla_{\theta} \mathbb{E}_{\phi \sim \mathcal{N}(\theta, \sigma^2 I)}\left[ \mathcal{I}(T,\phi)\right] 
    &= \nabla_{\theta} \int p_{\theta}(\phi) \mathcal{I}(T,\phi) d\phi \\
    &= \int p_{\theta}(\phi) \nabla_{\theta} \text{log} p_{\theta}(\phi) \mathcal{I}(T,\phi) d\phi \\
    &= \int p_{\theta}(\phi) \frac{(\phi - \theta)}{\sigma^2} \mathcal{I}(T,\phi) d \phi.
\end{align}

With change of variable $\phi=\theta+ \sigma \epsilon$, we have
\begin{align}
     \nabla_{\theta} \mathbb{E}_{\phi \sim \mathcal{N}(\theta, \sigma^2 I)}\left[ \mathcal{I}(T,\phi)\right] 
    &= \int p_{\mathbb{I}}(\epsilon) \mathcal{I}(T,\theta + \sigma \epsilon) \epsilon / \sigma d \epsilon\\
    &= \mathbb{E}_{\epsilon \sim \mathcal{N}(0,I)} \left[ \mathcal{I}(T,\theta + \sigma \epsilon) \epsilon / \sigma\right].
\end{align}

\section{Comparing standard benchmark protocols and different parameterizations of the control}
\label{app:extra_data}
In addition to the SA benchmark results in the main paper, in this appendix we compare the ML results to optimized parameterized standard bang-bang~\cite{Karzig2015,Yang2017} and super-adiabatic ramp-up-down protocols~\cite{Conlon2019}. We will also show results for ML optimizations of different parameterizations of the control $x_L(t)$. 

For a standard bang-bang protocol the velocity of the domain wall $v_L$ is at all times only allowed to take either the value $v_L=v_{\text{min}}$ or the value $v_L=v_{\text{max}}$, a switch between these two discrete velocities is called a bang. To search for the optimal bang-bang protocols we fix the minimum time between two consecutive bangs to be $\Delta t=0.1$ and perform a simulated annealing search similar to the one described in the main text. In this SA search we impose an additional constraint (compared to the free search in the main text) in which each update steps consist of changing $v_{\text{max}}$ to $v_{\text{min}}$ and vice versa to retain the bang-bang character of the protocols. We note that a version of this method was used in \cite{PhysRevLett.122.020601} to search for optimal bang-bang protocols in a spin system. We used $v_{\text{min}}=0$ similar to the bang-bang protocols in~\cite{Karzig2015} and we scanned different values of the maximum velocity $0<v_{\text{max}}<8v_{\text{avg}}$. This means that for some of the bang-bang protocols in regimes I and II $v_{\text{max}}\geq v_{\text{crit}}$ which was not the case for the protocols in ~\cite{Karzig2015}, however we observed that increasing $v_{\text{max}}$ leads to lower infidelities even when making $v_{\text{max}}$ bigger than the critical velocity. 

The ramp-up-down protocols are a family of super-adiabatic protocols in which the velocity is slowly built up from zero to some maximum velocity $v_{\text{max}}$ and then ramped down again to zero as given by

\begin{equation}
    v_L(t) = 
    \left\{
	\begin{array}{ll}
		v_{\text{max}}\frac{1-\cos{\omega t}}{2},  &  0 \leq t \leq \frac{\pi}{\omega}\\
		v_{\text{max}}, &  \frac{\pi}{\omega} \leq t \leq \frac{\pi}{\omega}+T \\
		v_{\text{max}}\frac{1-\cos({\omega t - \omega T})}{2}, & \frac{\pi}{\omega}+T \leq t \leq \frac{2\pi}{\omega}+T \\
		0, & \text{otherwise} \\
	\end{array}
\right.
\end{equation} in which $\omega$ is the parameter that determines how quickly the velocity is accelerated to $v_{\text{max}}$. To find the best ramp-up-down protocol in each regime we scanned over a large range of values of the free parameter $\omega$ ($v_{\text{max}}$ is fixed by the average velocity constraint).  

In Table ~\ref{tb:N50_benchmark} we summarize the best results (lowest infidelity) of these protocols. We compared these results to protocols obtained with the ML methods as described in the main text and see that in all regimes the ML protocols outperform the standard benchmark protocols significantly. The optimal bang-bang protocols obtained with the SA method in regimes II and IV are plotted in Fig. \ref{fig:gradient_and_bang}. It can be seen that the protocols start to approximate the optimal JMJ and super-adiabatic protocols obtained with the other methods. To be able to fully approximate the optimal JMJ protocols with a bang-bang protocol one needs to test even higher maximum velocities than we did here and also consider a negative minimum velocity $v_{\text{min}}<0$. 

\begin{table*}[h!]
  \centering
  \begin{tabular}{|c|c|c|c|c|c|c|} \hline
  Regime & linear & bang-bang & ramp-up-down & SA & DP & NES \\\hline
  I & 0.4738 & 0.3923  & 0.4817  & 0.3801 & 0.3546  & 0.3431 \\\hline
  II & 0.2236& 0.1713 & 0.2350 & 0.1637 & 0.1549 & 0.1514 \\\hline
  III & 0.0120 &  0.0067 & 0.0125  & 0.0062 & 0.0056 & 0.0071 \\\hline
  IV & 0.0077  & 0.0004 & 0.0045  & 0.0004  & 0.0005 & 0.0009 \\\hline
  \end{tabular}
  \caption{Results for the infidelity $\mathcal{I}(T)$ of different benchmark protocols (first three columns) in regime I-IV compared to results obtained with SA, AD and NES. For these simulations $(l,T)$ are $\{(4.32,12),(4.95,22),(0.48,8),(2.4,40)\}$, $N=110$ and the other parameters of the Majorana wire setup are the same as for figure \ref{fig:overview} in the main text.}
  \label{tb:N50_benchmark}
\end{table*}

We also tested the NES and DP optimization algorithms for different parameterizations of the control of the domain wall on a smaller system size $N=50$ before fixing the specific parameterizations as used and described in the main text for the big system $N=110$. We compared parameterizing the control by the position $x_L(t)$, the velocity $v_L(t)$ or a neural network $x_L(t)= \text{NN}_{\theta}(t)$ and show the results in Table ~\ref{tb:N50_AD_benchmark}. The optimization with NES gives the lowest infidelity when we parameterize by position in regimes I and II and by velocity in regimes III and IV. For the DP optimization the best infidelity optimizes a neural network in regimes I,III and IV and optimizes over the position in regime II. 

\begin{table*}[h!]
  \centering
  \begin{tabular}{|c|c|c|c|c|c|c|c|} \hline
  Regime & DP neural net & DP position & NES position & NES velocity & NES neural net \\\hline
  I & 0.3910  & 0.4159   & 0.3945 & 0.4008 & 0.4586\\\hline
  II & 0.1541 & 0.1531  & 0.1453 & 0.1509 & 0.1842 \\\hline
  III & 0.0064  & 0.0070    & 0.0120 & 0.0069 & 0.0084 \\\hline
  IV & 0.0001  & 0.0024     & 0.0148 &0.0004 & 0.0007\\\hline
  \end{tabular}
  \caption{Comparison of the lowest infidelity values between optimizations of AD and NES with respect to different parameterizations of the control (position $x_L(t)$, velocity $v_L(t)$ and neural net $x_L(t)= \text{NN}_{\theta}(t)$). Based on these testing values we determined which parameterizations to use for the optimizations in the main text. These simulations were run for a system size of $N=50$ and all the other parameters are the same as the parameters in Fig. \ref{fig:overview} in the main text.}
  \label{tb:N50_AD_benchmark}
\end{table*}

\section{\label{app:mlqcontrolcomp} Discussion on other methods for quantum control} 

We now briefly discuss the NES and DP methods in relation to traditional quantum control algorithms and RL methods. First we note that DP provides a general framework to automatically evaluate gradients of computer programs which can be integrated into traditional quantum control techniques such as the Krotov method~\cite{krotovGlobalMethodsOptimal1993},  Gradient-based-Pulse-Engineering\cite{khanejaOptimalControlCoupled2005}, and other gradient-based control techniques. In this context, a key advantage of DP is its ability to ease the implementation of gradient-based techniques, where, e.g., the incorporation of physical constraints, such as the restrictions in the amplitude of the controls, requires only adding the constraints to the main objective function. This avoids complex and potentially error-prone analytical calculations. In addition, the implementation of control protocols to widely different physical scenarios (e.g. free fermion or boson Hamiltonians or alternative simulation strategies) typically requires only the coding of the forward propagation of the simulation with minimal changes to the optimization subroutines. In addition, quantum optimal control algorithms based on DP can easily harness the acceleration afforded by graphics processing units (GPUs), which may speedup quantum control calculations by more than an order of magnitude~\cite{Leung2017}. This is most easily achieved using freely available open-source packages such as JAX~\cite{jax2018github}. 

Compared to RL and other gradient-free methods, the DP method used here remains a powerful method as long as the physical model is differentiable, as is the case in our work since the BdG algorithm \ref{app:gradient_analysis} used to compute the many-body infidelity is fully differentiable. DP and all gradient-based methods provide a strong optimization signal and typically converge faster than RL techniques, which often require high sample complexity for good convergence. In addition, the use of gradients avoids the  reward function design problem in RL \cite{matheron2019problem,dulacarnold2019challenges}. This allows us to efficiently extend DP to the many particle Hamiltonian required for the Majorana transport. 

On the other hand, NES exhibits multiple advantageous features. First, as a gradient-free black box optimization method, NES is the most flexible among the methods implemented in our work since it can be applied to controls that are discrete and continuous. Second, it can optimize non-differentiable control problems, as well as problems where the gradient estimation is numerically unstable in a way that makes the application of DP challenging. NES is also easily parallelizable across multiple processors for a fast collection of sufficient samples for optimization. Lastly, compared to RL,  we mention that NES can be directly applied to any desired objective function. This avoids the problem of the design of reward function commonly encountered in RL algorithms, which is necessary to avoid sparse learning signals that may hamper the efficiency of RL methods in quantum control.   

We also briefly mention that RL techniques such as the Watkins Q-learning algorithm \cite{10.5555/3312046} and policy gradient, tabular Q-learning, and deep Q-learning have recently been shown to be capable of achieving high performance in other quantum control setups~\cite{zhangWhenDoesReinforcement2019}. Ref.\cite{zhangWhenDoesReinforcement2019} provides a detailed comparison between tabular Q-learning, deep Q-learning, and policy gradient~\cite{10.5555/3312046}, and gradient descent and Krotov algorithms applied to the problem of preparing a target quantum state. Gleaned from their numerical experiments, the authors in Ref.~\cite{zhangWhenDoesReinforcement2019} conclude that the deep Q-learning and policy gradient algorithms tend to outperform  other RL and traditional control algorithms when the problem allows for discrete values of the control parameters, and when the problem is scaled up to large system sizes. 

Finally, we comment on well-established adiabatic shortcut methods~\cite{RevModPhys.91.045001}, in particular counter diabatic driving (CDD)  and Lewis-Riesenfeld invariants (LRI), in the context of the Majorana game problem. We note that an attempt to implement these in our setup encounters an array of problems. In CDD for example, the control protocols required a level of Hamiltonian control far beyond what would be experimentally feasible, where a full time-dependent and spatial control of a the problem Hamiltonian is required. Lastly, LRI requires finding a suitable invariant, which to the best our knowledge, is not known for the Majorana wire.

\section{Semiconducting nanowire spin-orbit coupled to s-wave superconductor}
\label{app:semiconductor}
The Kiteav chain model for topological p-wave superconductivity \cite{Kitaev2001} used for the simulations in the main text can be  realized effectively in a semiconducting nanowire proximity coupled to a conventional s-wave superconductor in a Zeeman field \cite{PhysRevLett.100.096407,PhysRevLett.105.177002,PhysRevLett.105.077001}. In this appendix we will first introduce the proximity coupled model and discuss in which limits the Kitaev chain toy model can be obtained. Then we show how the critical velocity $v_{\text{crit}}$ and resonance frequency can be obtained such that we can define the same Majorana motion regimes as in the main text. Finally we show numerical optimization results obtained with AD and NES in each of those regimes for two different sets of parameters of the proximity coupled model (with one set outside the Kitaev chain limit). 

\subsection{Hamiltonian and Energy dispersion}

The Hamiltonian for the proximity coupled semiconducting nanowire (NW) for periodic boundary conditions, a homogeneous potential $V(x)=V$ and chemical potential $\mu(x)=\mu$ in the continuum momentum representation is given by

\begin{equation}
    \mathcal{H}_{\text{NW}} = \frac{1}{2}\int dk \Psi^\dagger_k H^{\text{BdG}}(k)\Psi_k 
\end{equation} with $\Psi^\dagger_k = [c^\dagger_{k\uparrow} c^\dagger_{k\downarrow} c_{-k\uparrow} c_{-k\downarrow}]$ and 

\begin{equation}\label{eq:bdg_swave}
    H^{\text{BdG}}(k) = (\frac{k^2}{2m} + V- \mu + B\sigma_x + \alpha k\sigma_y)\tau_z + \Delta \sigma_y\tau_x
\end{equation} in which $\sigma_i$ acts on the spin-degrees of freedom and $\tau_i$ on the particle-hole space. In here we have an external magnetic field $B$ in the x-direction, spin-orbit coupling with strength $\alpha$ in the y-direction and a standard s-wave superconducting term with gap $\Delta$. Note the main difference with the Kiteav chain in Eq. \ref{eq:Hamcont} is the spin-degrees of freedom $s=\{\uparrow\downarrow\}$ which result in Zeeman splitting, spin-orbit interaction and s-wave superconductivity instead of spin-less p-wave superconductivity.
    
The energy dispersion can be found by solving the characteristic equation for $H^{\text{BdG}}$ Eq. \ref{eq:bdg_swave} which leads to \begin{equation}
    \epsilon_{\pm}^2(k) = \xi^2 + B^2 + \alpha^2k^2 + \abs{\Delta}^2 \pm 2\sqrt{B^2\abs{\Delta}^2 + \xi^2B^2 + \alpha^2k^2\xi^2}
\end{equation} in which we have defined $\xi(k)=\frac{k^2}{2m} + V- \mu$. In this four-band model various parameter choices lead to a gapped dispersion with in some phases Majorana zero modes, see for a complete discussion e.g. \cite{PhysRevLett.105.177002}. Moreover, in two specific limits (large Zeeman field $B\gg\alpha, \Delta$ or for large spin-orbit coupling $\alpha\gg B, \Delta$)  this proximity coupled NW model can be projected onto the lower bands (reduces to two similar blocks) resulting in the Kitaev chain model with p-wave superconducting gap $\Delta'=\frac{\alpha\Delta}{B}$ (large $B$) or $\Delta'=\frac{\Delta}{m\alpha}$ (large $\alpha$) respectively \cite{Scheurer2013,Rieder2012}.

\subsection{Critical Velocity and Resonance Frequency}
\begin{figure*}[t!]
    \centering
    \includegraphics[width=1.0\linewidth]{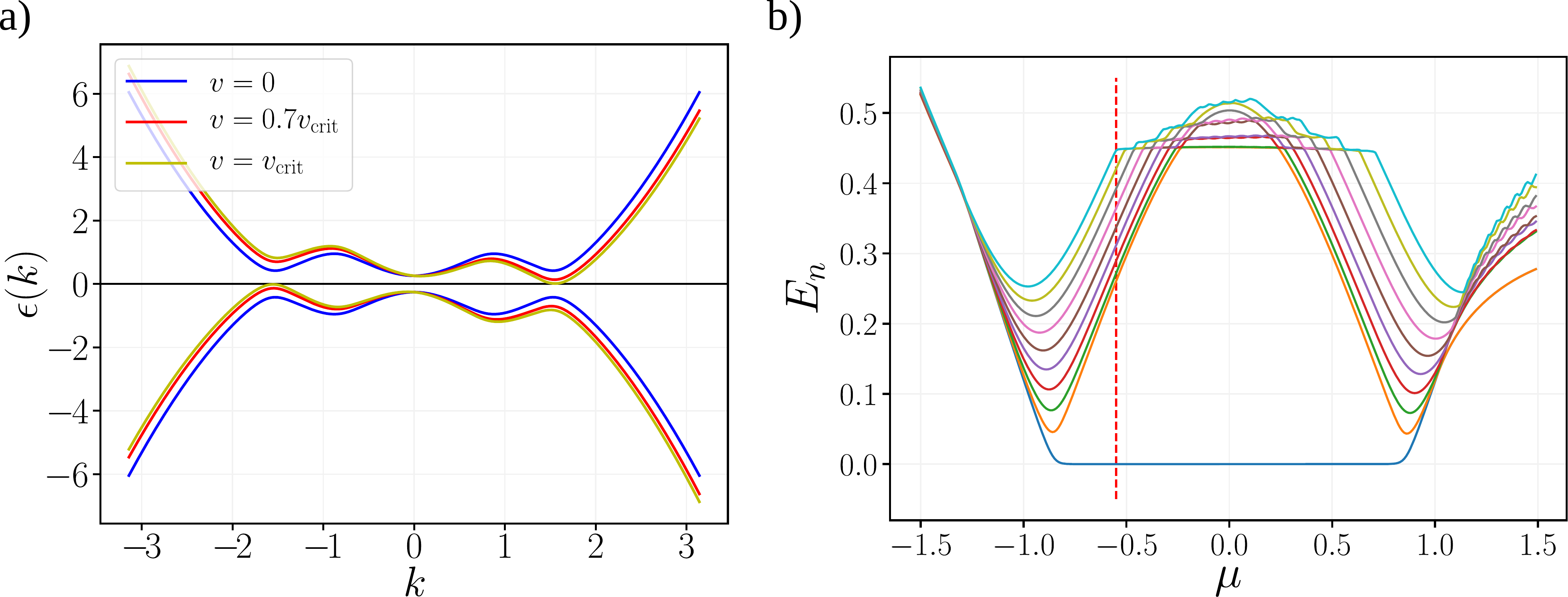}
    \caption{a) Energy dispersion of the Hamiltonian Eq. \ref{eq:BdG_Ham_mov} in the moving frame for three different velocities. The velocity tilts the dispersion and the gap closes when $v=v_{\text{crit}}$. The model parameters used for this dispersion are $B = \alpha = 1.0, \Delta=0.5$ and $\mu=-0.55, V=0.0$. b) Excitation energy eigenvalues $E_n$ of the NW model on a lattice for open boundary conditions as a function of the chemical potential $\mu$. The red dashed line indicates $\mu=-0.55$. The other parameters are the same as for panel a). We get a zero mode (blue line) but also some Andreev bound states (e.g. orange line) that come down when we bring $\mu$ closer to the bottom of the band.}  
    \label{fig:critvel}
\end{figure*}

To bring the NW in Eq. \ref{eq:bdg_swave} into the moving frame we apply the same techniques as for the Kitaev chain discussed earlier in App. \ref{app:majorana_motion}, i.e. we apply the time-dependent translation operator $U(t)=e^{-ik\int_0^tv(t')dt'}$ to rotate the Hamiltonian Eq. \ref{eq:bdg_swave}, which results in the moving frame Hamiltonian \begin{equation}\label{eq:BdG_Ham_mov}
    H_v^{\text{BdG}}(k) = \left(\frac{k^2}{2m} + V- \mu + B\sigma_x + \alpha k\sigma_y\right)\tau_z + \Delta \sigma_y\tau_x + vk\mathbb{1}_4. 
\end{equation} Like for the Kitaev chain this transformation gives an additional contribution $vk$ to the energy dispersion that tilts it as shown in figure \ref{fig:critvel}. In the regimes when the system is originally gapped the gap closes due to this tilt when $v\equiv v_{\text{crit}}=\frac{E_{\text{gap}}}{k_F}$. For most parameters finding the general expression for the energy gap and critical velocity is too complicated and we need to obtain it numerically. However things simplify in the Kitaev chain limit for $B\gg \alpha, \Delta$ which gives $v_{\text{crit}} \approx \frac{\alpha\Delta}{B}$ and when $\alpha\gg B, \Delta$ we obtain $v_{\text{crit}} \approx \frac{\Delta}{m\alpha}$ \cite{Scheurer2013}.  

The resonance frequency in this model is like for the Kitaev chain defined as the being equal to the gap $\omega_{\text{res}}=E_{\text{gap}}$. With these associations for the critical velocity and resonance frequency the definition of the Majorana motion regimes originally in terms of parameters of the Kitaev chain can be extended to include the proximity coupled model as well; regime I is defined to be for $v_{\text{avg}}>v_{\text{crit}}$, regime II close but below the critical velocity, regime III low velocity and short time-scales with respect to the energy gap and finally the (super)-adiabatic regime IV for long times and low velocities with respect to the gap.

\subsection{Optimization results}

By putting the NW model Eq. \ref{eq:bdg_swave} on a lattice, opening up the boundary condition and imposing the smooth potential profile in Eq. \ref{eq:Potentialprofile} we acquire the same control of the position of the Majorana zero mode as for the Kitaev chain in the main text. In Fig. \ref{fig:critvel} b) we show an example of the excitation energies for a set of parameters outside the Kitaev chain limit in which we have a Majorana zero mode and also Andreev bound states. 

\begin{figure*}[t!]
    \centering
    \includegraphics[width=1.0\linewidth]{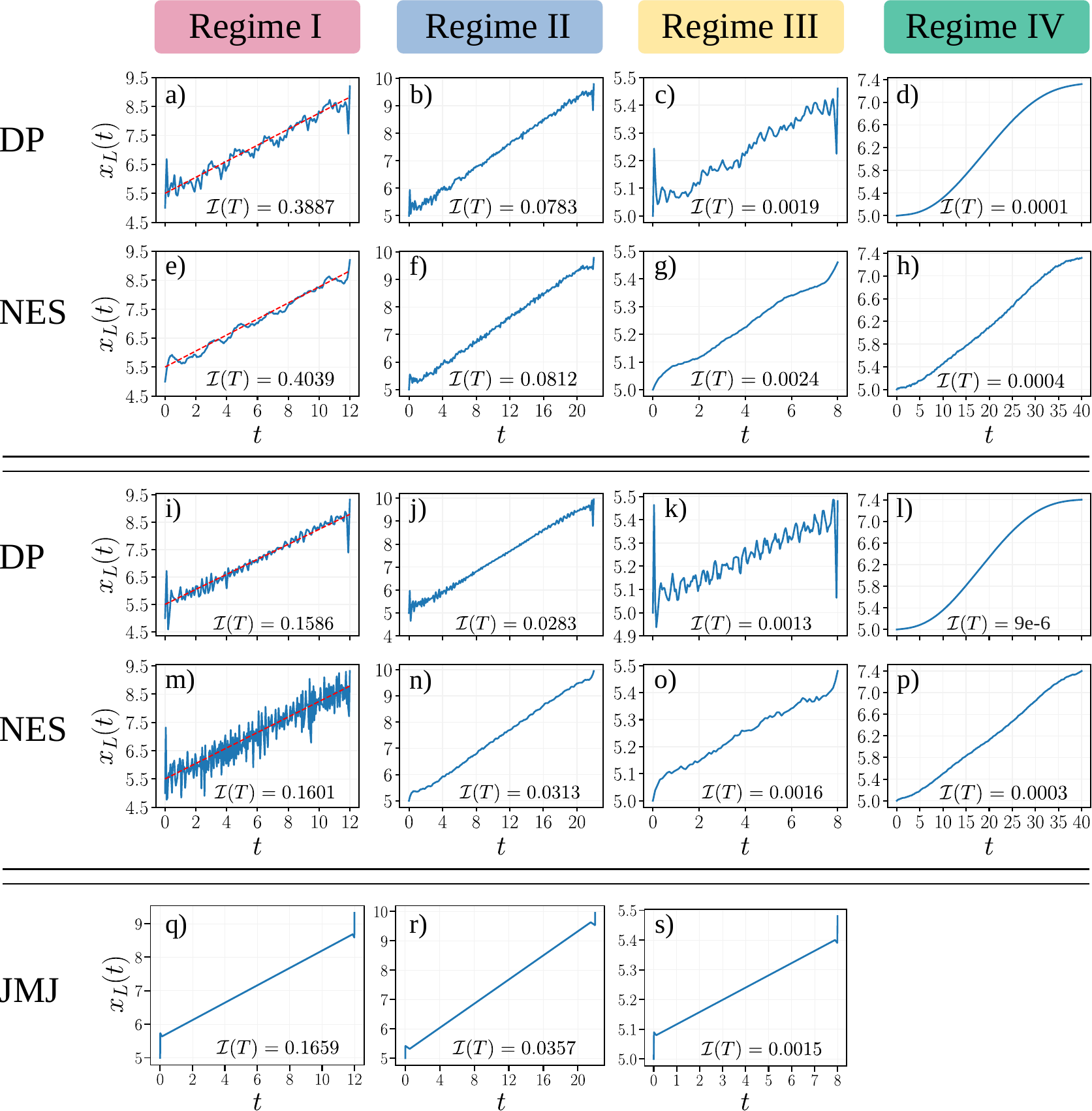}
    \caption{Overview of the optimal controls in the four different Majorana motion regimes (I-IV) obtained after optimization with Differentiable Programming (DP), Natural Evolutionary Strategies (NES) and the parameterized JMJ protocols for the proximity coupled NW model as given by Eq. \ref{eq:bdg_swave}. In all panels we plot the domain wall position $x_L(t)$ as a function of time starting from $x_L(0)=x_A=5.0$. The red dotted line in regime I has a slope equal to the critical velocity $v_{\text{crit}}\approx0.27$ for panels a/e and $v_{\text{crit}}\approx0.28$ for panels i/m. The obtained strategies are similar to the strategies obtained for the Kitaev chain in the main text with pulse-like motion at the beginning and end in regimes I-II-III bookening smoother motion in the middle and in regime IV a smooth adiabatic protocol. In the simulations for panels a-h we set $N=100$, $\mu=0.0$, $w=1$, $\Delta=0.8$, $B=2$, $\alpha=0.8$, $V_{\text{height}}=30.1$, and $\sigma=1$ and for panels i-s we set $N=100$, $\mu=-0.55$, $w=1$, $\Delta=0.5$, $B=1$, $\alpha=1.0$, $V_{\text{height}}=30.1$, and $\sigma=1$. The constraint parameters $(l,T)$ are given by  $\{(4.32,12),(4.95,22),(0.48,8),(2.4,40)\}$ for regimes I,II,III, and IV, respectively. }  
    \label{fig:resultsNW}
\end{figure*}

We then optimize the transport of the Majoranas in this NW model with the same AD and NES techniques as used before.  We show the results for these simulations for the same parameters as in Fig. \ref{fig:critvel} from Differentiable Programming and Natural Evolutionary Strategies in Fig. \ref{fig:resultsNW} (i-p) and from JMJ protocols in Fig. \ref{fig:resultsNW} (q-s). For completeness in Fig. \ref{fig:resultsNW} (a-h) we include results obtained by choosing parameters that approximate the Kitaev chain limit (high $B-$field). 

It can be seen that for these two sets of parameters we obtained qualitatively similar strategies for Majorana transport in the proximity-coupled NW as for the simpler Kitaev chain model as shown in Fig. \ref{fig:overview} in the main text. That is, in regions I-II-III we get the JMJ strategy with a pulse-like motion at the beginning and end of the protocol and an on average constant motion with a velocity $v_{\text{crit}}$ in the middle of the protocol. For the parameters outside the Kitaev limit, the initial and final dressed jumps are comprised of three parts: a forward jump, a backward jump and another forward jump. For these parameters we observe a much smoother protocol in regime II compared to the Kitaev chain. In regime IV for both sets of parameters we again recover the expected smooth adiabatic protocols. 

To show that the JMJ-strategy (together with its analysis discussed in the main text) is also a robust strategy for the proximity coupled NW model we scanned again over the different parameters $\Delta x_{\text{forward}}$ and  $\Delta x_{\text{back}}$ outside the Kitaev limit. The optimal JMJ protocols obtained in this way together with the infidelity values are shown in Fig. \ref{fig:resultsNW} (q-s). It can be seen that the infidelity values for the dressed JMJ protocol are competitive to the protocols obtained with DP and NES. Compared to the Kitaev chain results (panel (d) in Fig. \ref{fig:majorana_game_setup}) it can be seen that for this proximity coupled NW model the dressing of the jumps is slightly less pronounced.

\section{Parameters and interpolation of the optimized JMJ model strategy}
\label{app:model}

In this appendix we give the parameters of the optimized JMJ strategies in regimes I,II, and III as shown in Fig.~\ref{fig:majorana_game_setup}~d) in the main text and verify how close these strategies are to the protocols obtained with the ML methods. In the main text we observed that dressed JMJ protocols with the lowest infidelities have a linear relation between $\Delta x_{\text{forward}}$ and $\Delta x_{\text{backward}}$ as Fig.~\ref{fig:simple_model} shows. To attain the best JMJ protocols in regime I-III, we scan through $\Delta x_{\text{forward}}$ and $\Delta x_{\text{back}}$ with respect to the linear relation in each regime and pick the one which gives the overall lowest infidelity. The parameters of the resulting JMJ strategies are provided in Table ~\ref{tb:best_JBMBJ} and the strategies are shown in Fig. \ref{fig:majorana_game_setup}.
\begin{table*}[h!]
  \centering
  \begin{tabular}{|c|c|c|c|c|c|} \hline
  Regime & $\Delta x_{\text{forward}}$ & $\Delta x_{\text{back}}$ & $\Delta t_{\text{back}}$ & Infidelity \\\hline
  I & 7.992  & 7.5060  & 0.05 & 0.3575\\\hline
  II & 1.9093  & 1.5496  & 0.31 & 0.1589\\\hline
  III & 0.5482  & 0.4571   & 0.33 & 0.0057\\\hline
  \end{tabular}
  \caption{Best dressed JMJ model strategies parameters and their corresponding infidelities obtained from scans over $\Delta x_{\text{forward}}$, $\Delta x_{\text{back}}$ and $\Delta t_{\text{back}}$. The corresponding profiles are shown in Fig. \ref{fig:majorana_game_setup} in the main paper.}
  \label{tb:best_JBMBJ}
\end{table*}

\begin{figure}[h!]
    \centering
    \includegraphics[width=1.0\linewidth]{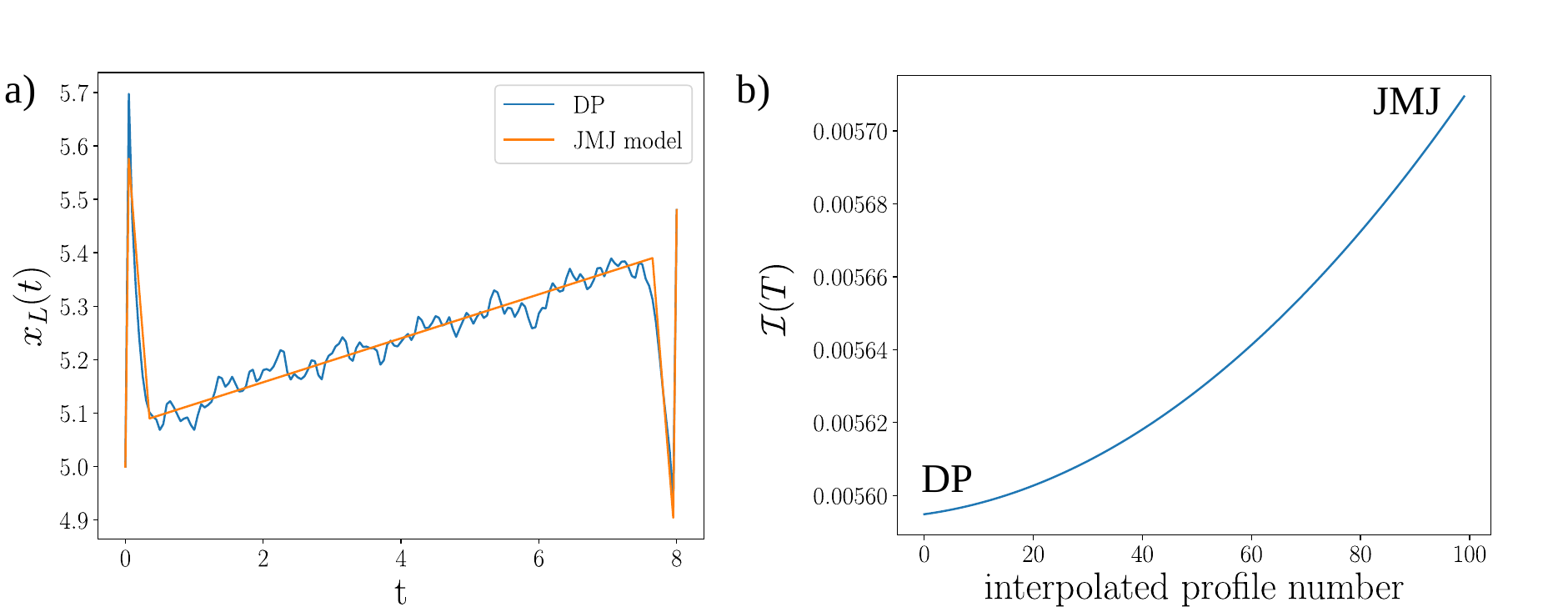}
    \caption{a) The DP optimized majorana motion strategy in regime III $(T=8,l=0.48)$ on top of the optimized (scans over the free parameters) JMJ model strategy. The optimal JMJ model strategy captures the main features (jumps) of the DP strategy well. b) Infidelity $\mathcal{I}(T)$ obtained for 100 linearly interpolated profiles between the DP strategy and JMJ model strategy as shown in panel a). We observe a convex decreasing function towards the more flexible DP result. }
    \label{fig:interpolation}
\end{figure}

We linearly interpolate these best JMJ model strategies to the  machine learning protocols in the main paper (Fig. \ref{fig:overview}) to see whether there are additional protocols that are better in terms of infidelity. The results for this in regime III with the DP strategy are shown in Fig. \ref{fig:interpolation} from which we can see that the DP strategy is slightly better than the JMJ (due to more free parameters) and that there are no other local minima in between them. For the other regimes and interpolations from the NES strategy we obtained similar results.  

\section{Extra Results on Robustness of the JMJ strategy with respect to Interactions and Disorder}

\begin{figure*}[t!]
    \centering
    \includegraphics[width=1.0\linewidth]{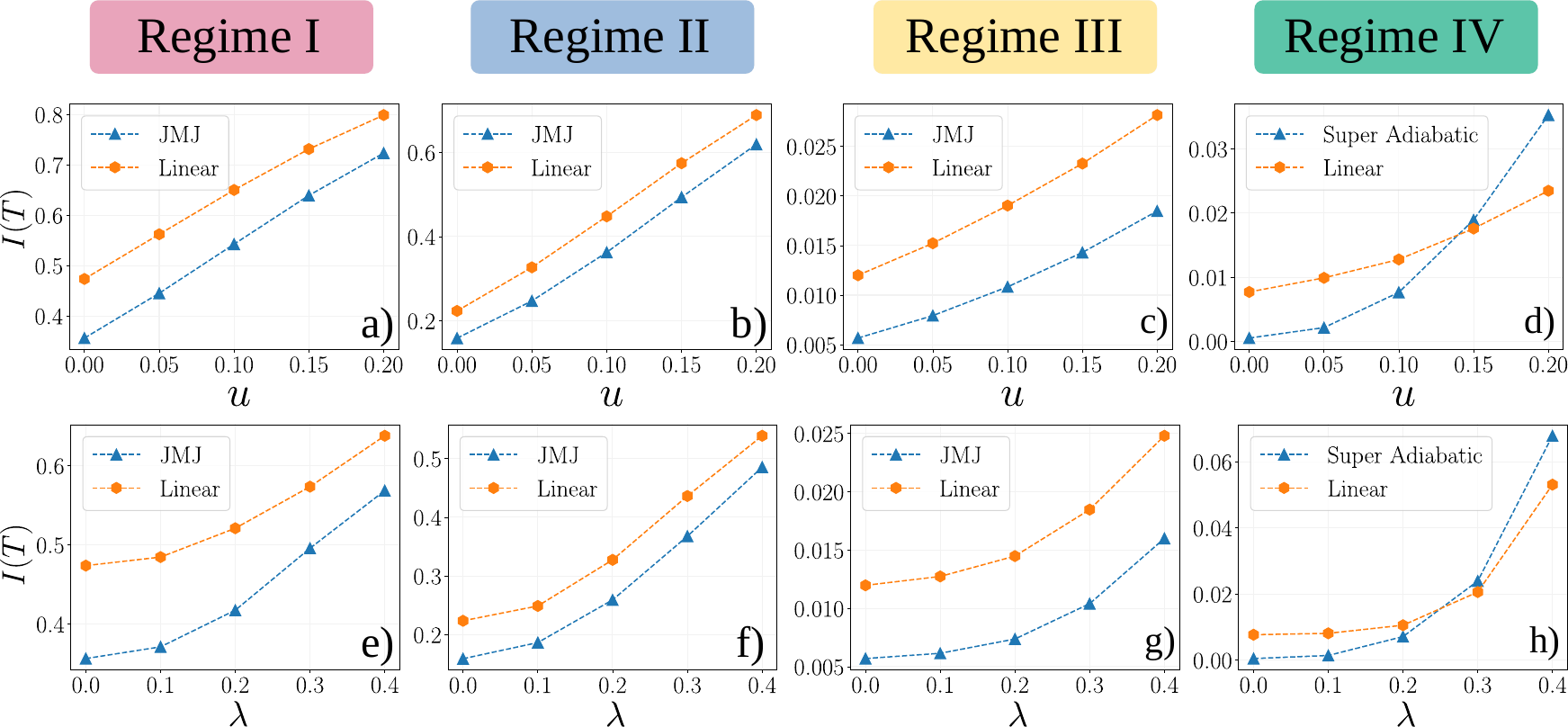}
    \caption{Overview of simulation results of running the optimal protocols obtained in the clean non-interacting system in a system with interactions (top row) or disorder (bottom row) in all four regimes. In all regimes the infidelity increases with disorder strength $\lambda$ and also with interaction strength $u$. The parameters for these simulations were the same as in Fig. \ref{fig:overview} in the main text and the disorder averaging was done over 500 disorder realizations.}  
    \label{fig:dis_int}
\end{figure*}

In Fig. \ref{fig:dis_int} we show simulation results when running the optimal protocols obtained in the clean system in a system with disorder and separately interactions. In summary we see that in regimes I, II and III the ML protocols in the clean non-interacting system outperform a naive linear benchmark protocol. On the other hand, in regime IV after large enough interaction strengths or disorder strengths the linear protocols outperform the super-adiabatic protocols found before in the literature. Moreover, the infidelity values get worse by more than an order of magnitude in this regime which emphasises the importance of looking for alternative strategies that can be more efficient than the smooth super-adiabatic protocols.

The observation that the infidelity values generally get worse with increasing interaction strength can be explained, at least partially, due to a lowering of effective critical velocity. This comes about in two main ways. Firstly on a mean-field level we know that including repulsive onsite interactions results in a lower topological energy gap $E_{\text{gap}}$ (see e.g \cite{Stoudenmire2011}).  This would result in, as shown in \cite{Coopmans2020}, the resonance frequency $\omega_{\text{res}}$ becoming lower which means that the resonance time-scale, and consequently regime IV, get pushed to longer timescales $T$.

Another feature that may be at play here is the fact that interactions, on a mean-field level, typically induce non-uniformity in the effective couplings (found for example by running some suitable Hartree-Fock-Bogoliubov minimisation \cite{Ring2004}). The result in this case would be similar to the scenario encountered in the disordered situations  where moving the wall over changing landscape amounts to a time dependent perturbation (see e.g. \cite{Karzig2013}). For future studies it would be interesting to investigate if one can use NES or DP to improve these values further by optimizing in the presence of disorder or interactions. 

\label{app:int_dis}

\section{Jump infidelity costs with system parameters}
\label{app:JinfCost}

In this appendix we examine the infidelity cost of a single jump as a function of the important physical lengths scales $\lambda_F= 2 \pi /k_F$ and $\xi = 1/(m \Delta)$. Our numerical test suggest a near Gaussian drop off in the infidelity as a function of jump distance: $O_\delta \sim \exp (- \delta^2/s^2)$. Generally speaking, the larger the value of $s$, the larger one can jump. 

\begin{figure*}[h!]
    \centering
    \includegraphics[width=1.0\linewidth]{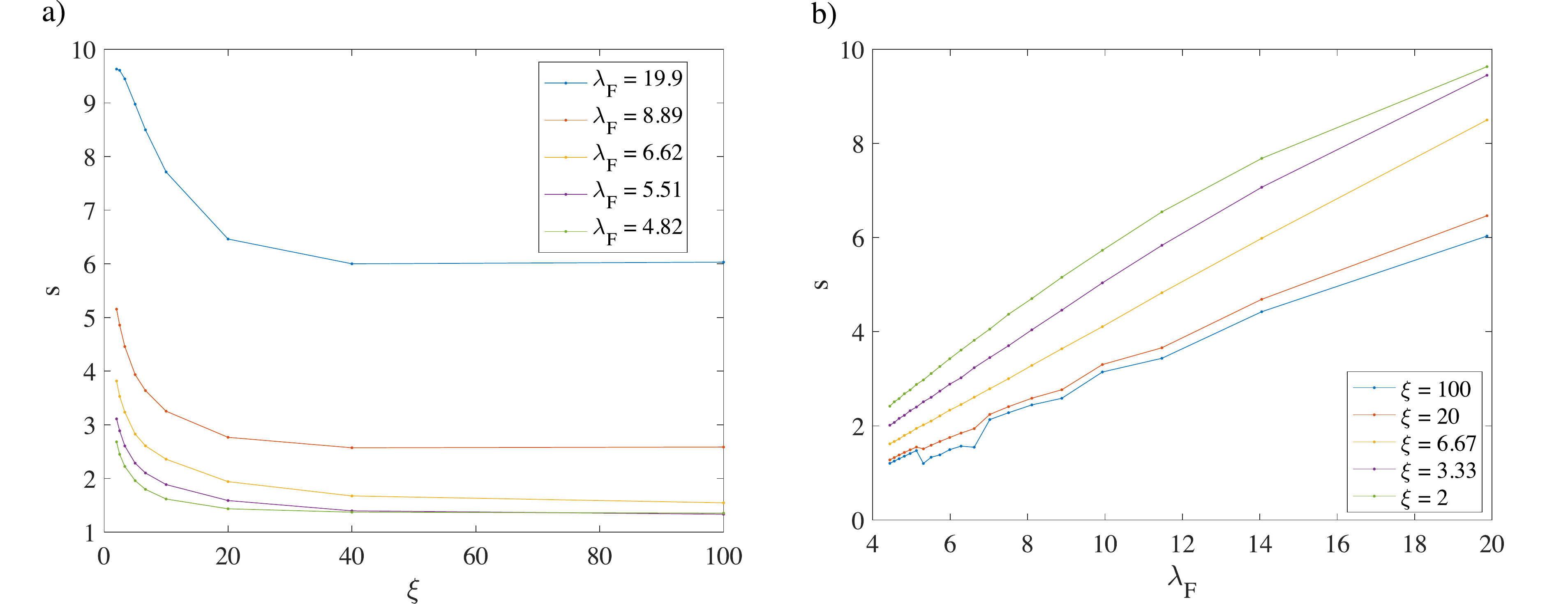}
    \caption{In the p-wave model the $O_\delta\sim \exp (- \delta^2/s^2)$ have an interesting relationship to the two key length scales. The fit parameter $s$ is reduced by a factor of about $1/2$  when the coherence length $\xi=1/ (m \Delta) $ is made very large. On the other hand its value tends to increase linearly in proportion to $\lambda_F = 2\pi/ k_F$. The topological gap $E_{\text{gap}} \sim \Delta k_F = 2 \pi / (m \xi \lambda_F)$ is however inversely proportional to both length scales.   }  
    \label{fig:s_versus}
\end{figure*}

In figure \ref{fig:s_versus} we plot the value of $s$ that we obtain from a numerical fitting the $O_\delta$ drop-off.  We see that the value of the superconducting parameter $\Delta$ (and hence $\xi$) do play a role: the value of $s$ tends to go down, but not dramatically so, as $\Delta$ gets smaller (and $\xi$ gets larger). Perhaps more importantly we see the that $s$ parameter is linearly related to $\lambda_F$.  This means the jump size can in principle get bigger if one tunes the chemical potential towards the bottom of the single particle valence band. 

\end{document}